\documentclass[aps,prl,reprint,nofootinbib,preprintnumbers,floatfix]{revtex4-1}

\pdfoutput=1

\usepackage[bookmarks=false,hyperfootnotes=false]{hyperref}
\usepackage{graphicx}
\usepackage{amsmath}
\usepackage{subfig}
\usepackage{xspace}
\usepackage{slashed}

\usepackage{bbm}

\newcommand{\ecf}[2]{e_{#1}^{(#2)}}

\newcommand{\Dobs}[2]{D_{#1}^{(#2)}}

\newcommand{\Cobs}[2]{C_{#1}^{(#2)}}

\DeclareRobustCommand{\Fig}[1]{Fig.~\ref{#1}}

\DeclareRobustCommand{\Ref}[1]{Ref.~\cite{#1}}

\newcommand{\Nsub}[2]{\tau_{#1}^{(#2)}}

\newcommand{\pythia}[1]{\textsc{Pythia\xspace #1}}

\newcommand{\fastjet}[1]{\textsc{FastJet\xspace #1}}

\newcommand{\herwigpp}[1]{\textsc{Herwig++\xspace #1}}

\begin{document}

\preprint{MIT--CTP 4595}

\title{Building a Better Boosted Top Tagger}

\author{Andrew J. Larkoski}
\email{larkoski@mit.edu}
\author{Ian Moult}
\email{ianmoult@mit.edu}
\author{Duff Neill}
\email{dneill@mit.edu}

\affiliation{Center for Theoretical Physics, Massachusetts Institute of Technology, Cambridge, MA 02139, USA}

\begin{abstract}
Distinguishing hadronically decaying boosted top quarks from massive QCD jets is an important challenge at the Large Hadron Collider.  In this paper we use the power counting method to study jet substructure observables designed for top tagging, and gain insight into their performance. We introduce a powerful new family of discriminants formed from the energy correlation functions which outperform the widely used $N$-subjettiness. These observables take a highly non-trivial form, demonstrating the importance of a systematic approach to their construction.
\end{abstract}

\maketitle

Boosted top quarks arising from decays of heavy resonances occur in many new physics models motivated by the hierarchy problem, and are an important and well studied signal at the Large Hadron Collider (LHC) \cite{Cheng:2005as,Barger:2006hm,Agashe:2006hk,Lillie:2007yh,Baur:2007ck,Baur:2008uv,Pilot:2013bla,Fleischmann:2013woa}. The problem of discriminating these boosted top quarks, which exhibit a three-pronged substructure, from the background of massive QCD jets has received considerable attention in the jet substructure literature \cite{Kaplan:2008ie,Thaler:2008ju,Almeida:2008yp,Almeida:2008tp,Plehn:2009rk,Plehn:2010st,Almeida:2010pa,Thaler:2010tr,Thaler:2011gf,Jankowiak:2011qa,Soper:2012pb,Larkoski:2013eya}. A variety of techniques have been tested in new physics searches at the LHC \cite{CMS:2011bqa,Aad:2012dpa,Aad:2012raa,Chatrchyan:2012ku,TheATLAScollaboration:2013qia,Aad:2013gja,CMS:2014aka}, with one of the most effective discriminants being the $N$-subjettiness ratio observable $\Nsub{3,2}{\beta}$ \cite{Thaler:2010tr,Thaler:2011gf}.

The construction of efficient discriminating observables is typically guided by Monte Carlo simulations. While Monte Carlos play an essential role at the LHC, this approach introduces dependence on non-perturbative tunings, and often leaves unclear whether an optimal observable has been identified. Only recently has a program for analytic understanding of the simplest substructure observables been developed \cite{Catani:1991bd,Seymour:1997kj,Ellis:2010rwa,Chien:2010kc,Feige:2012vc,Dasgupta:2012hg,Chien:2012ur,Krohn:2012fg,Waalewijn:2012sv,Field:2012rw,Jouttenus:2013hs,Dasgupta:2013via,Dasgupta:2013ihk,Larkoski:2013paa,Larkoski:2014uqa,Larkoski:2014tva,Larkoski:2014pca,Procura:2014cba,usD2}, with more complex variables, such as those required for boosted top tagging, far beyond the current level of calculability. An important question then is whether optimal discriminating observables can be predicted without requiring a complete analytic calculation. In \Ref{Larkoski:2014gra} it was shown that power counting techniques, which incorporate the parametric predictions of soft and collinear QCD emissions, can be used to make robust predictions about the behavior of jet substructure variables, and provide a systematic approach to identify optimal discriminating observables. This was explicitly demonstrated by studying observables for boosted $Z$ boson identification. Power counting was also used in \cite{Walsh:2011fz} to study constraints on jet substructure algorithms.

In this paper we extend the power counting analysis of \Ref{Larkoski:2014gra} to the case of boosted top discrimination. By analyzing the phase space formed by the simultaneous measurement of three energy correlation functions, $ (\ecf{2}{\alpha}, \ecf{3}{\beta},\ecf{4}{\gamma} )$, we will show that power counting arguments alone identify a powerful family of discriminating variables, $D_3^{(\alpha,\beta,\gamma)}$. These variables outperform the $N$-subjettiness observable $\Nsub{3,2}{\beta}$ in both \pythia{8} and \herwigpp. Their complicated form emphasizes the need for a systematic approach to their construction, which is provided by power counting. 

The $n$-point energy correlation functions are \cite{Larkoski:2013eya}
\begin{align}\label{eq:ecf_def}
\ecf{n}{\beta} =\frac{1}{p_{TJ}^n}\sum_{i_1<\cdots <i_n \in J}   \left ( \prod_{a=1}^n p_{T_{i_a}}  \right )   \left( \prod_{b=1}^{n-1} \prod_{c=b+1}^{n} R_{{i_b} {i_c}} \right )^\beta \,,
\end{align}
where $p_{TJ}$ is the transverse momentum of the jet, and the angular exponent satisfies $\beta>0$ for infrared and collinear (IRC) safety. The boost-invariant angle $R_{ij}^2 = (\phi_i-\phi_j)^2+(y_i-y_j)^2$, where $\phi$ is the azimuth and $y$ is the rapidity. \Ref{Larkoski:2013eya} proposed the observables $\Cobs{2}{\beta}$ and $\Cobs{3}{\beta}$ for identifying two- and three-prong jets, respectively, which are defined as
\begin{align}
\Cobs{2}{\beta}=\frac{  \ecf{3}{\beta}   }{ ( \ecf{2}{\beta}  )^2  } \ , \qquad
\Cobs{3}{\beta}=\frac{  \ecf{4}{\beta}   \ecf{2}{\beta}   }{ ( \ecf{3}{\beta}  )^2  } \ .
\end{align}
Recently, using power counting methods, \Ref{Larkoski:2014gra} defined a new observable for two-prong jet discrimination,
\begin{equation}
\Dobs{2}{\beta}=\frac{  \ecf{3}{\beta}   }{ ( \ecf{2}{\beta}  )^3  }  \ .
\end{equation}
$\Dobs{2}{\beta}$ was shown to be more robust to a mass cut and contamination in the jet than $\Cobs{2}{\beta}$, as well as providing improved discrimination power.
Applying the same techniques to the problem of boosted top tagging leads to the definition of $D_3^{(\alpha,\beta,\gamma)}$ as we will discuss. This variable exhibits considerably improved discrimination power compared with $\Cobs{3}{\beta}$.

At the high energy scales probed at the LHC, QCD is approximately conformal, with jets being dominated by soft and collinear radiation whose intrinsic energy and angular scales are determined by measurements on the jet. Power counting identifies the parametric scalings associated with soft and collinear radiation, which dominate the observable, allowing for robust predictions about the behavior of substructure variables. Since the predictions rely on parametric scalings, they must be reproduced by any Monte Carlo generator.

\begin{figure*}[t]
\centering
\subfloat[]{\label{fig:triple_NINJA}
\includegraphics[width=5cm]{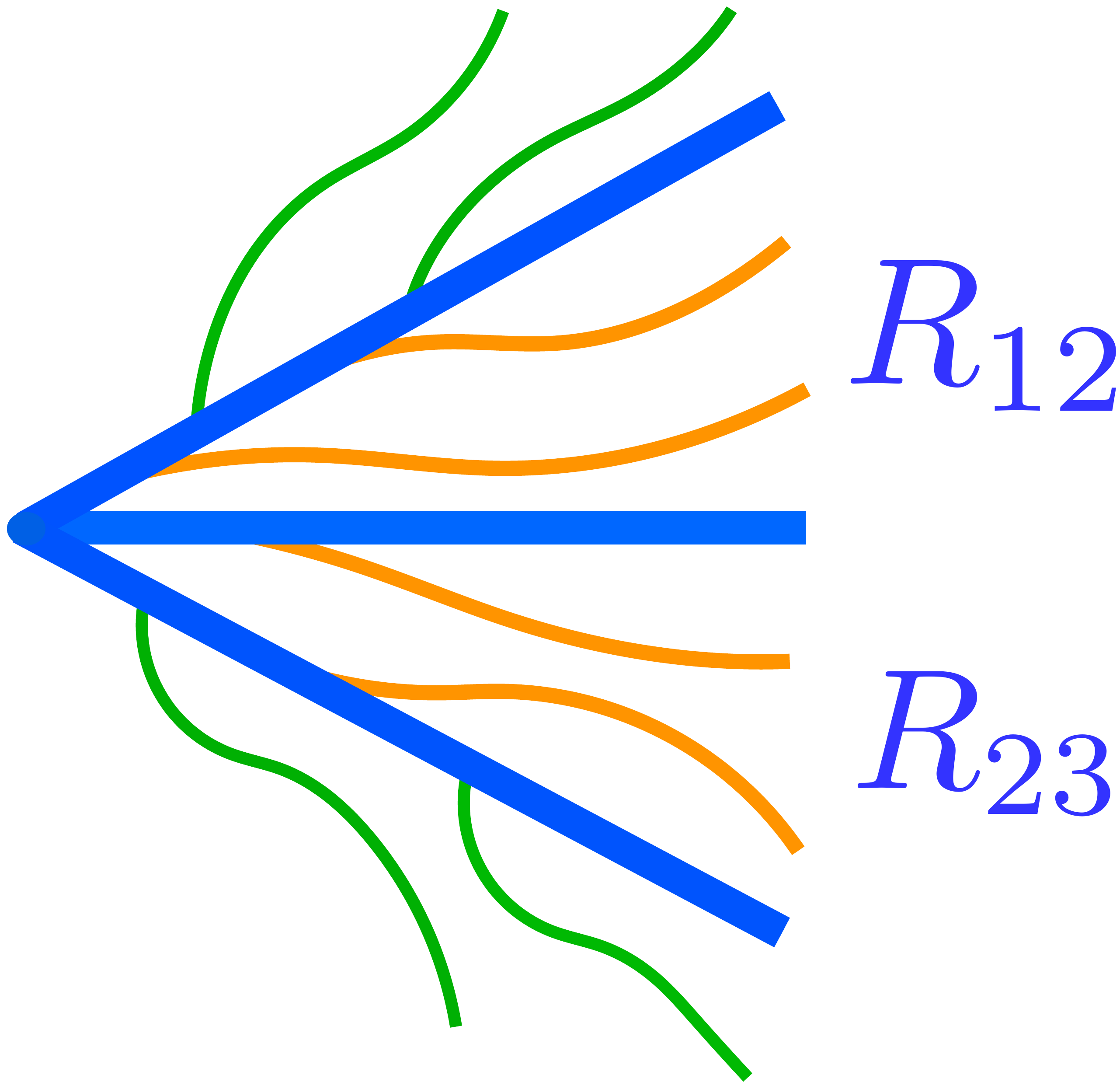}    
} \ \ 
\subfloat[]{\label{fig:p_NINJA}
\includegraphics[width=5cm]{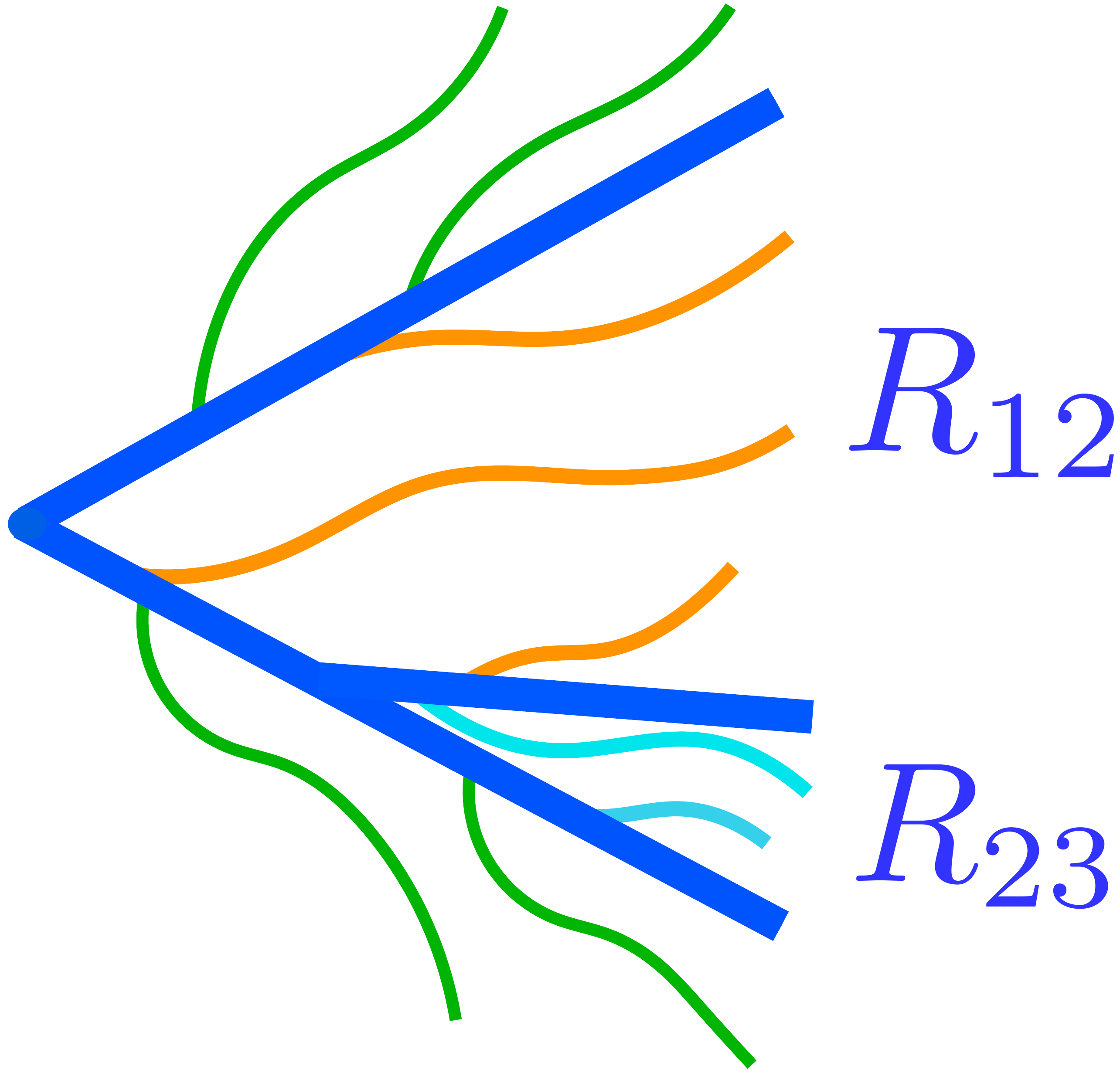} 
}\ \ 
\subfloat[]{\label{fig:s_NINJA}
\includegraphics[width=6.3cm]{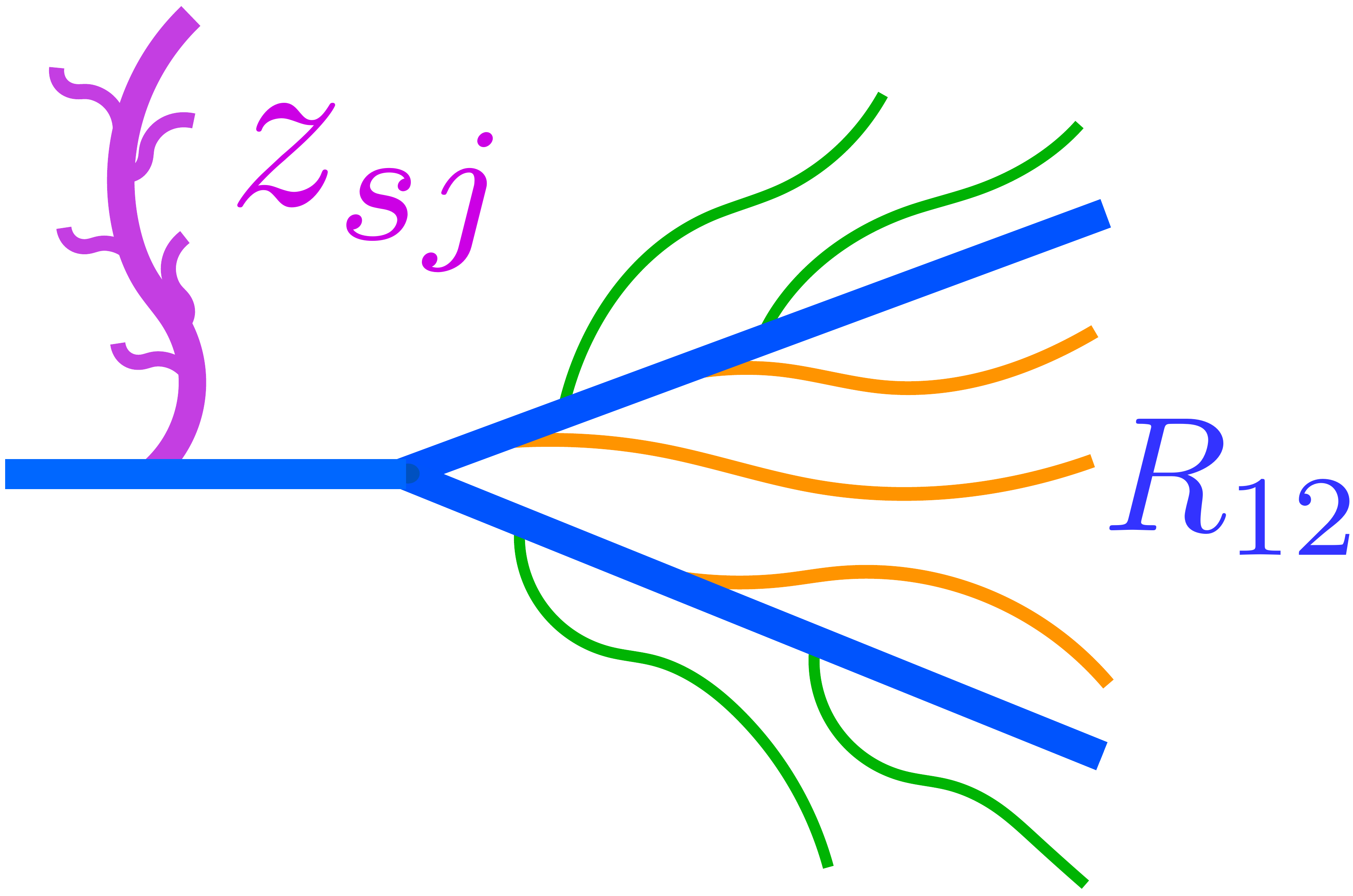}
}
\caption{Three prong jet configurations (a) Triple Splitting, (b) Strongly Ordered Splitting, and (c) Soft Emission, required to understand the $ (\ecf{2}{\alpha}, \ecf{3}{\beta},\ecf{4}{\gamma} )$ phase space. The structure of dominant energy flow for each configuration is shown in blue, as well as soft radiation (green), and radiation from the dipoles (orange, light blue). 
}
\label{fig:triple_NINJAS}
\end{figure*}

To identify the optimal top tagger formed from the energy correlation functions, we apply a power counting analysis to the $( \ecf{2}{\alpha} ,\ecf{3}{\beta} ,\ecf{4}{\gamma} ) $ phase space. The simultaneous measurement of these three observables allows for the resolution of up to three subjets within a jet, as is required for discriminating boosted top quarks from massive QCD jets.  By studying the region of phase space occupied by three prong jets, in particular the scaling of its boundaries, the optimal ratio observable can be determined.

The jet configurations with three resolved subjets that are required to understand the boundaries of the phase space are shown in \Fig{fig:triple_NINJAS}. Each configuration, and its relevant soft and collinear modes, will be discussed in turn.  For this analysis, we will assume that the angular exponents $\alpha$, $\beta$, and $\gamma$ are all ${\cal O}(1)$ and not parametrically different.  Details of the motivation for a power counting analysis and its application to hadronically-decaying boosted $W/Z/H$ bosons are provided in \Ref{Larkoski:2014gra}.  We will provide some details of the phase space constraints for the configuration in \Fig{fig:triple_NINJA} to illustrate the procedure, while for the other configurations, we will simply state the results.

\begin{itemize}
\item {\bf a): Triple Splitting} 

\Fig{fig:triple_NINJA} shows a three-pronged jet with no hierarchies between the angles or energies of the jets: $R_{12} \sim R_{23}\sim R_{13}\ll 1$, and all subjets carry an ${\cal O}(1)$ fraction of the jet $p_T$.  Then, the dominant contributions to the two- and three-point energy correlation functions are determined by the angles between the hard subjets:
\begin{align}
\ecf{2}{\alpha} &\sim R_{12}^\alpha+R_{23}^\alpha+R_{13}^\alpha \,, \nonumber \\
\ecf{3}{\beta} &\sim \left(R_{12}R_{23}R_{13}\right)^\beta \,.
\end{align}
Therefore, this configuration populates the region of phase space above $\ecf{3}{\beta} \sim ( {\ecf{2}{\alpha}} )^{3\beta/\alpha}$.

The four-point energy correlation receives dominant contributions from the radiation off of the hard subjets.  This radiation consists of collinear emissions at characteristic angle $R_{cc}$, soft radiation at large angles (shown in green), and radiation from the subjet dipoles (called ``collinear-soft'' radiation \cite{Bauer:2011uc}, shown in orange).  Soft radiation has a $p_T$ fraction $z_s$ and is emitted at ${\cal O}(1)$ angles while collinear-soft radiation has $p_T$ fraction $z_{cs}$ and is emitted at angles comparable to the separation of the subjets, denoted by $R_{cs}\sim R_{12}$.  Then, the dominant contributions to the four-point energy correlation function are
\begin{equation}
\hspace{0.75cm}\ecf{4}{\gamma}\sim \left(R_{cc}^{\gamma}R_{12}^{2\gamma}+z_{s}+z_{cs}R_{12}^{3\gamma}\right)\left(R_{12}R_{23}R_{13}\right)^\gamma \,.
\end{equation}
With the assumption that the three-prong structure is well-defined, $R_{cc}\ll R_{12}$ and $z_s\ll z_{cs}\ll 1$, in this region, the four-point energy correlation function satisfies:
\begin{equation}
\frac{ \ecf{4}{\gamma} }{   ({\ecf{3}{\beta}} )^{2\gamma/\beta}   }\ll 1\,.
 \end{equation}

\item {\bf b): Strongly Ordered Splitting}

\Fig{fig:p_NINJA} shows a three-pronged jet with hierarchical opening angles: $R_{23}\ll R_{12}$, but with all subjets carrying ${\cal O}(1)$ of the jet $p_T$. In addition to soft, collinear, and collinear-soft modes, modes which we call collinear-collinear-soft, emitted from the dipole of the second branching (shown in light blue) are required to describe the radiation in the jet.

This configuration populates the region of the $(\ecf{2}{\alpha} , \ecf{3}{\beta} )$ plane defined by   $\ecf{3}{\beta} \ll ({\ecf{2}{\alpha}} )^{3\beta/\alpha}$. The parametric scaling of the upper boundary for this region of phase space is found to be
\begin{equation}
\frac{   \ecf{4}{\gamma}   \left ({\ecf{2}{\alpha}}\right)^{3\gamma/\alpha}  }{     \left( \ecf{3}{\beta}\right )^{3\gamma/\beta}     }\ll 1\,,
\end{equation}
which agrees with the scaling of the triple splitting configuration for $\ecf{3}{\beta} \sim ( {\ecf{2}{\alpha}} )^{3\beta/\alpha}$.

\item {\bf c): Soft Emission} 

\Fig{fig:s_NINJA} shows a three-pronged jet in which one subjet has a $p_T$ fraction which is parametrically smaller than the other two: $z_{sj} \ll 1$. Soft, collinear, and collinear-soft modes are required for the description of the two subjets with small opening angle. Additional collinear-soft modes are required for the description of the soft subjet, with characteristic $p_T$ fraction, $z_{sj}$.

Different hierarchies between the opening angle, $R_{12}$, and the soft jet $p_T$ fraction, $z_{sj}$, identify distinct scalings.
In the case that $z_{sj}\gg R_{12}^\alpha$, the upper boundary of the phase space region is given by
\begin{align}
\frac{   \ecf{4}{\gamma}   \left (\ecf{2}{\alpha}\right)^{2\gamma/\beta-1}  }{      \left (\ecf{3}{\beta}\right )^{2\gamma/\beta}    } \ll 1\,,
\end{align}
while for $z_{sj}\ll R_{12}^\alpha$, we have
\begin{equation}
 \frac{ \ecf{4}{\gamma} (\ecf{2}{\alpha})^{2\beta/\alpha -\gamma/\alpha}  }{ ( \ecf{3}{\beta})^2  } \ll 1\,.
\end{equation}
The power counting analysis in this soft jet region also suggests that optimal behavior for the observable requires $\beta, \gamma \lesssim 1$, where regions with distinct scalings are well separated.

\end{itemize}

In addition to these three configurations, there are other phase space regions to consider; for example, there is also a three pronged structure corresponding to two soft subjets coming off of a hard central core.  However, all other phase space regions are fully contained within the three listed above, and so do not provide additional constraints.  Finally, we leave for future work the investigation of the structure of the phase space when the jets have one or two prongs, since detecting the three-pronged structures is most important for the boosted top analysis.

\begin{figure}[t]
\includegraphics[width=6.2cm]{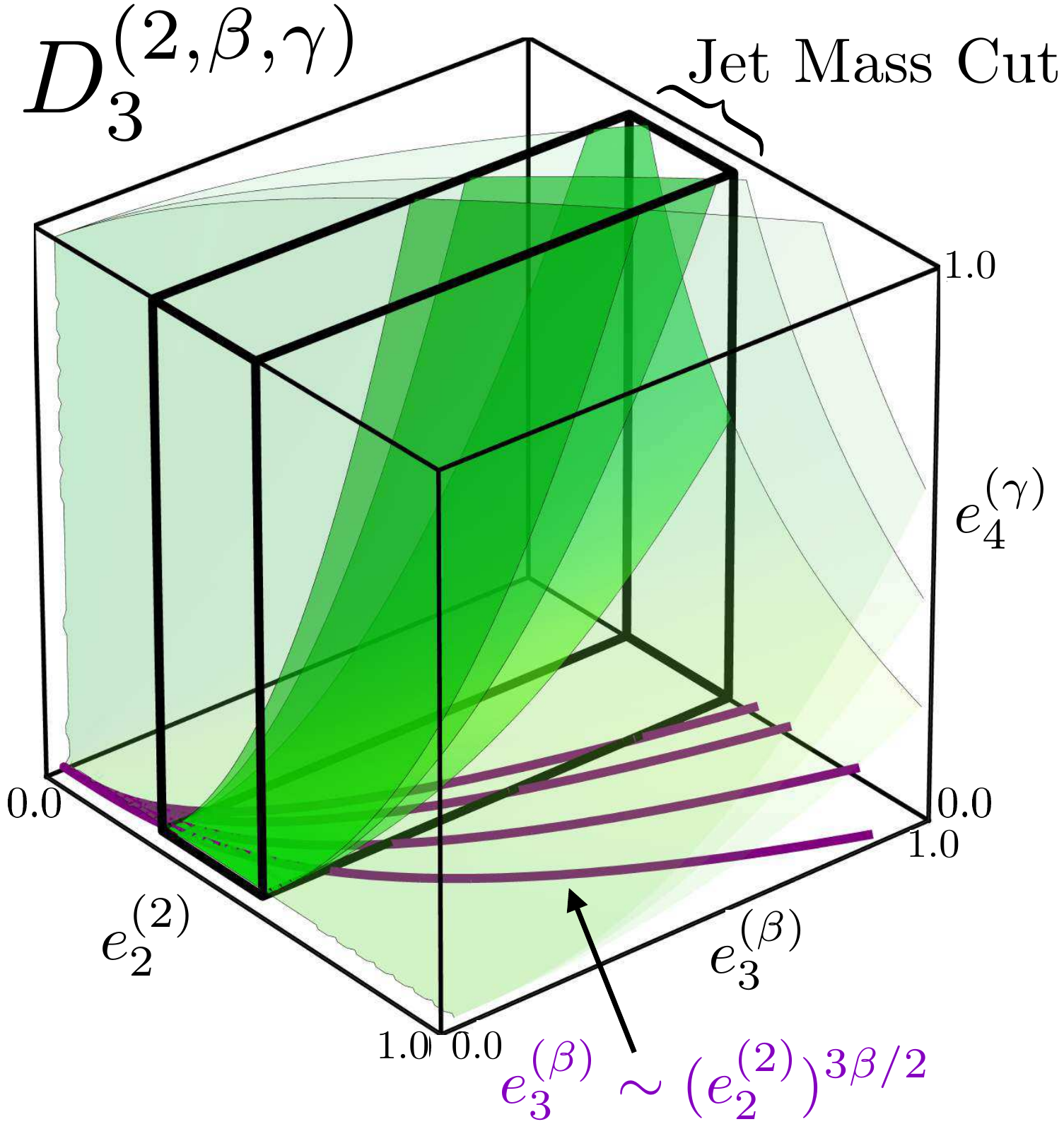}    
\caption{Phase space defined by the energy correlation functions, with contours of $D_3^{(2,\beta,\gamma)}$, and showing the effect of a jet mass cut.
}
\label{fig:phase_space}
\end{figure}

To construct a variable which discriminates between boosted top quarks and QCD jets, we use the fact that top quark jets will primarily populate the three prong region of phase space, while QCD jets will primarily populate the one or two prong regions. We therefore wish to find a variable which is small in the three prong regions of phase space, and which becomes large outside these regions. The variable must interpolate between the three different scalings we have found, and therefore we consider the sum,
\begin{align}\label{eq:D3_def}
&D_3^{(\alpha,\beta,\gamma)}=\\ \nonumber
&  \frac{   \ecf{4}{\gamma}   \left ({\ecf{2}{\alpha}}\right)^{\frac{3\gamma}{\alpha}}  }{     \left( \ecf{3}{\beta}\right )^{\frac{3\gamma}{\beta}}     } +x \frac{   \ecf{4}{\gamma}   \left (\ecf{2}{\alpha}\right)^{\frac{2\gamma}{\beta}-1}  }{      \left (\ecf{3}{\beta}\right )^{\frac{2\gamma}{\beta}}    }   +y \frac{   \ecf{4}{\gamma}   \left (\ecf{2}{\alpha}\right)^{\frac{2\beta}{\alpha}-\frac{\gamma}{\alpha}}  }{      \left (\ecf{3}{\beta}\right )^{2}    }\,,
\end{align}
where $x,y$ are as of yet undetermined constants.  We choose this summed form because it is the simplest combination that smoothly interpolates between the different three-pronged phase space regions.  This interpolation is necessary to robustly define the boundary of the three-prong region of phase space, where signal lives, from the rest of the phase space, where backgrounds live.  For compactness in the text, we will often write $D_3$, omitting the angular exponents. The power counting of the constants is determined by demanding that in the transition region, $\ecf{3}{\beta} \sim  (\ecf{2}{\alpha} )^{3\beta/\alpha}$, each term in the sum has the same power counting.  Using the fact that $\ecf{2}{\alpha}$ and the jet mass are related on the $\ecf{3}{\beta} \sim  (\ecf{2}{\alpha} )^{3\beta/\alpha}$ boundary, and assuming a tight cut on the jet mass in a window of the top quark mass, we find
\begin{equation}
x=\kappa_1 \left (\frac{(p^{\text{cut}}_T)^2}{m_{\text{top}}^2}\right )^{\left (\frac{\alpha \gamma}{\beta}-\frac{\alpha}{2} \right)},\quad y=\kappa_2   \left(\frac{(p^{\text{cut}}_T)^2}{m_{\text{top}}^2}\right )^{\left (\frac{5\gamma}{2}-2\beta \right)}\,.
\end{equation}
Since only the scaling of $x$, $y$ are determined by the power counting, an $\mathcal{O}(1)$ tuning of their values can be performed, and is represented by the variables $\kappa_1, \kappa_2\sim 1$. Here $p^{\text{cut}}_T$ is a proxy for the average jet $p_T$, which is dictated by the imposed cuts due to the steeply falling $p_T$ spectrum. Contours of $D_3$ in the $( \ecf{2}{2} ,\ecf{3}{\beta} ,\ecf{4}{\gamma} ) $ phase space are shown in  \Fig{fig:phase_space}.

\begin{figure}[t]
\includegraphics[width=6.5cm]{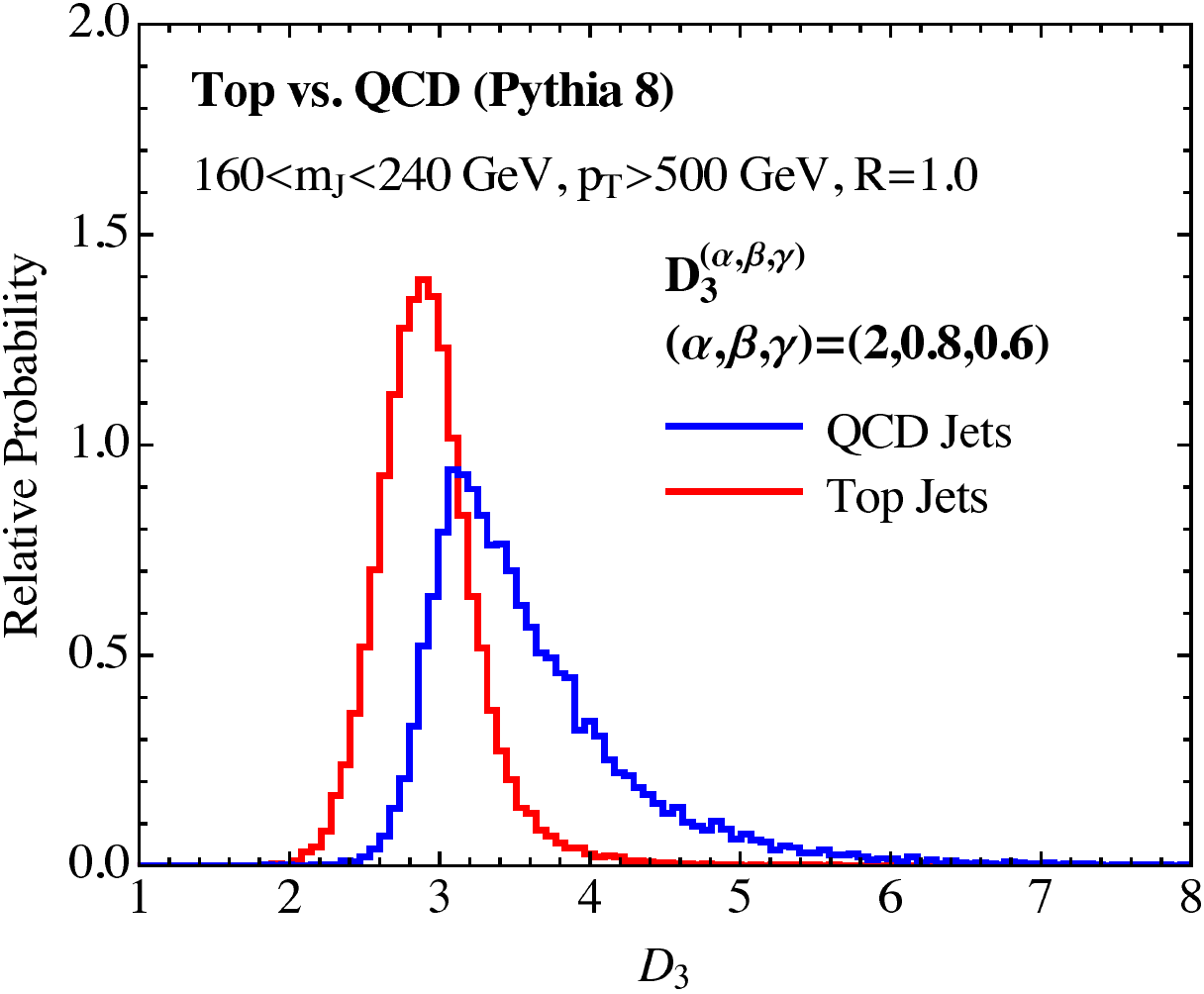}
\caption{Signal and background distributions for $D_3$ for the \pythia{8} samples.  Here, the parameters of $D_3$ are: $\alpha=2, \beta=0.8, \gamma=0.6$, and $x=5$, $y=0.35$.}
\label{fig:D3_distr}
\end{figure}

For $\alpha=2$, a cut on the jet mass gives a simple restriction on the phase space, due to the relation 
\begin{equation}
\ecf{2}{2}\sim \frac{m_J^2}{p_{TJ}^2}\,.
\end{equation}
In this case, a narrow cut on the jet mass effectively reduces the three dimensional $( \ecf{2}{2} ,\ecf{3}{\beta} ,\ecf{4}{\gamma} ) $ phase space to the two dimensional $(\ecf{3}{\beta} ,\ecf{4}{\gamma} ) $ phase space, as shown in \Fig{fig:phase_space}.  Away from $\alpha=2$, the constraint from the jet mass cut slices out a region of the phase space with complicated $ \ecf{2}{\beta}$ dependence, and is expected to reduce the discriminating power of the variable \cite{Larkoski:2014gra}. It would potentially be interesting to investigate the behavior of $D_3$ for $\alpha \neq 2$, however, this is beyond scope of this paper. Because of these considerations, we will restrict our attention to the variable $D_3^{(2,\beta,\gamma)}$.

\begin{figure*}[t]
\centering
\subfloat[]{
\includegraphics[width=7cm]{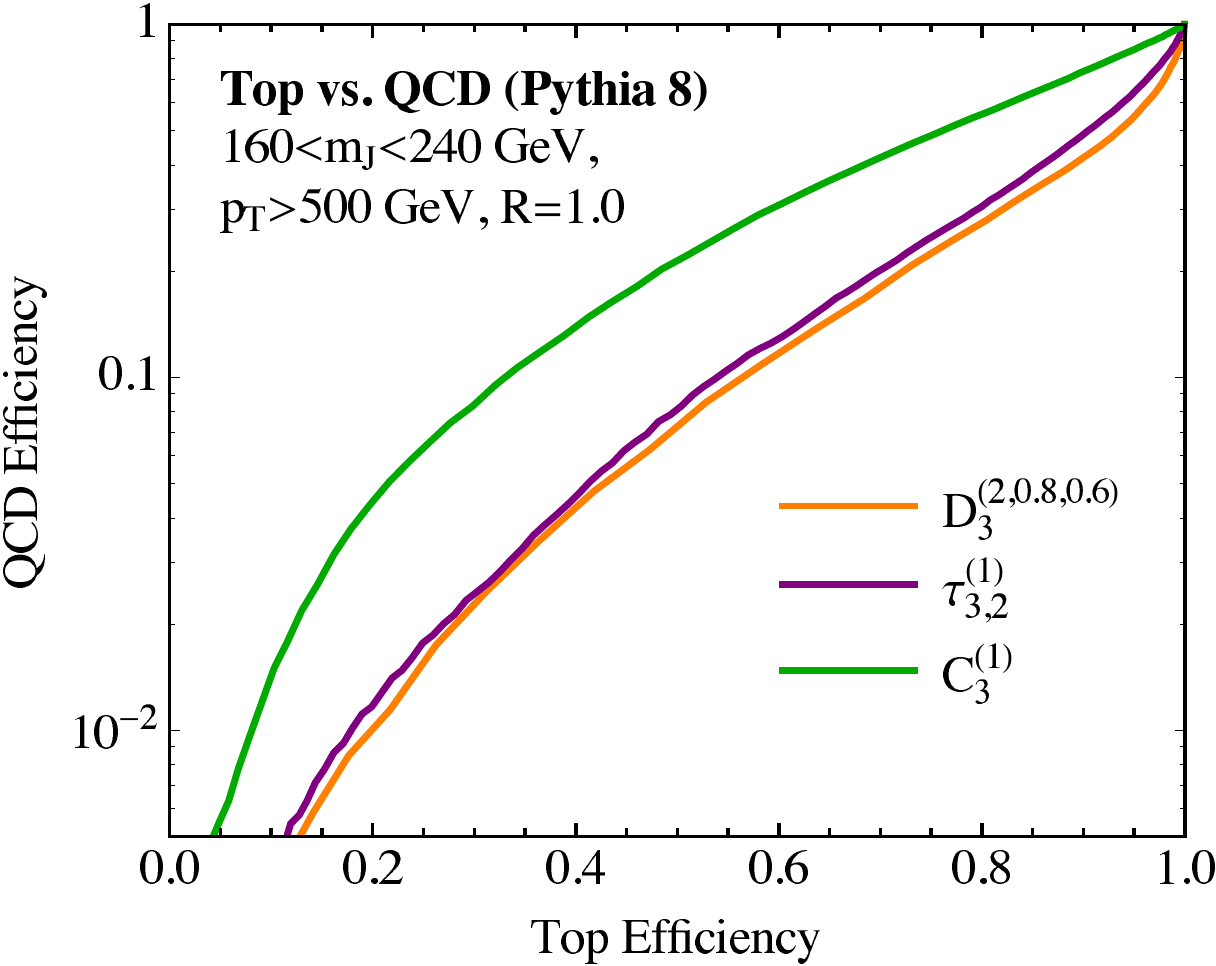}
}\qquad\qquad\qquad
\subfloat[]{
\includegraphics[width=7cm]{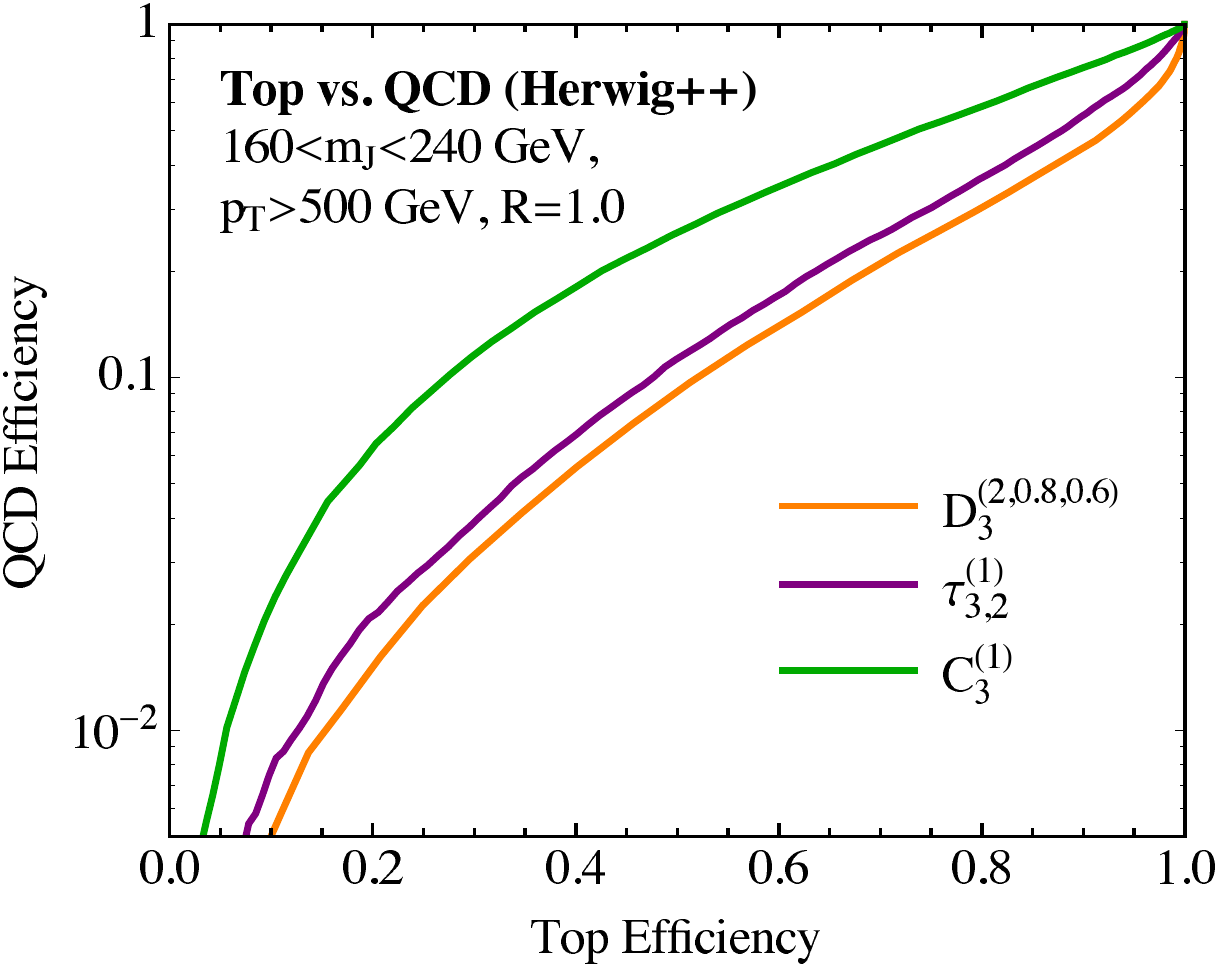}
}
\caption{Signal vs.~Background efficiency curves comparing $C_3^{(1)}$, $D_3^{(2,0.8,0.6)}$, and $\tau_{3,2}^{(1)}$ from \pythia{8} (left) and \herwigpp (right) samples.}
\label{fig:D3_ROC}
\end{figure*}

To test the discriminating power of $D_3$ we simulated $pp\to $ dijets and $pp\to t\bar t$ events at the 8 TeV LHC, with both tops decaying hadronically. Events were generated and showered with \pythia{8.183} \cite{Sjostrand:2006za,Sjostrand:2007gs} or \herwigpp{2.6.3} \cite{Marchesini:1991ch,Corcella:2000bw,Corcella:2002jc,Bahr:2008pv}. Fat jets with $R=1$ and $p_T>500$ GeV were clustered in \fastjet{3.0.3} \cite{Cacciari:2011ma} with anti-$k_T$ \cite{Cacciari:2008gp}, using the Winner Take All (WTA) recombination scheme \cite{Bertolini:2013iqa,Larkoski:2014uqa,Larkoski:2014bia,Salambroadening}. The energy correlation functions and $N$-subjettiness ratio observables were calculated using the \texttt{EnergyCorrelator} and \texttt{Nsubjettiness} \fastjet{contrib}s \cite{Cacciari:2011ma,fjcontrib}.

\Fig{fig:D3_distr} shows signal and background distributions for the variable $D_3$, as measured in \pythia{8}, which exhibit good signal/background separation. We have used angular exponents $\alpha=2, \beta=0.8, \gamma=0.6$, and $x=5$, $y=0.35$. No thorough optimization over these parameters has been explored, with $\alpha = 2$ to provide interpretation of $\ecf{2}{2}$ as the mass and the other values motivated by power counting and improving discrimination power. It is envisioned that optimization could be explored for particular situations, for example, for different $p_T$ bins or in the presence of pile-up.

In \Fig{fig:D3_ROC} we compare the signal vs. background efficiency (ROC) curves of $D_3$, defined with the same parameters as used in \Fig{fig:D3_distr}, with those of $\Cobs{3}{\beta}$ and $\Nsub{3,2}{\beta}$ for $\beta=1$, for which these variables are known to exhibit optimal performance \cite{Thaler:2010tr,Thaler:2011gf,Larkoski:2013eya}. $D_3$ exhibits the best performance over the entire range of signal efficiencies in both \pythia{8} and \herwigpp, with noticeable improvement in discrimination power at higher signal efficiencies. The effect of the mass cut is not included in the efficiencies. Power counting alone does not predict the relative performance of $D_3^{(\alpha,\beta,\gamma)}$ and $\Nsub{3,2}{\beta}$, which is determined by $\mathcal{O}(1)$ numbers. However, the power counting analysis shows that the energy correlation functions $( \ecf{2}{2} ,\ecf{3}{\beta} ,\ecf{4}{\gamma} ) $ provide a well-defined parametric separation of the phase space, which is not true for $N$-subjettiness \cite{Larkoski:2014gra}. This could be the explanation for the improved discrimination power of $D_3$.

An unambiguous prediction of power counting is that $D_3^{(\alpha,\beta,\gamma)}$ should provide much better discrimination power than $\Cobs{3}{\beta}$, which does not respect the different scalings of the phase space. This fact is evident from the signal versus background curve, with  $D_3^{(\alpha,\beta,\gamma)}$ significantly better than $\Cobs{3}{\beta}$ at all efficiencies. In \Ref{Larkoski:2013eya}, it was conjectured that the poor discrimination of $\Cobs{3}{\beta}$ was due to a proliferation of angular factors in the higher point energy correlation functions. However, we have shown that this is not the case, and that the use of power counting techniques can play an important role in constructing powerful jet substructure observables.

The observable $D_3$ identifies the parametric scalings in the energy correlation function phase space using an understanding of the behavior of QCD. Each term in the observable is associated with a physical configuration of subjets, and therefore is highly flexible, facilitating tuning for improved behavior under the addition of pile-up radiation, or in conjunction with grooming techniques. We envision that this level of flexibility can be utilized to optimize boosted top tagging much as is done with a multivariate analysis, but using a variable whose scalings are robust predictions of QCD.

We thank Jesse Thaler and Iain Stewart for helpful discussions. IM thanks Matthew Low for help with Monte Carlo issues. This work is supported by the U.S. Department of Energy (DOE) under cooperative research agreements DE-FG02-05ER-41360, and DE-SC0011090. D.N. is also supported by an MIT Pappalardo Fellowship. IM is also supported by NSERC of Canada.

\bibliography{toptag}

\begin{thebibliography}{61}%
\makeatletter
\providecommand \@ifxundefined [1]{%
 \@ifx{#1\undefined}
}%
\providecommand \@ifnum [1]{%
 \ifnum #1\expandafter \@firstoftwo
 \else \expandafter \@secondoftwo
 \fi
}%
\providecommand \@ifx [1]{%
 \ifx #1\expandafter \@firstoftwo
 \else \expandafter \@secondoftwo
 \fi
}%
\providecommand \natexlab [1]{#1}%
\providecommand \enquote  [1]{``#1''}%
\providecommand \bibnamefont  [1]{#1}%
\providecommand \bibfnamefont [1]{#1}%
\providecommand \citenamefont [1]{#1}%
\providecommand \href@noop [0]{\@secondoftwo}%
\providecommand \href [0]{\begingroup \@sanitize@url \@href}%
\providecommand \@href[1]{\@@startlink{#1}\@@href}%
\providecommand \@@href[1]{\endgroup#1\@@endlink}%
\providecommand \@sanitize@url [0]{\catcode `\\12\catcode `\$12\catcode
  `\&12\catcode `\#12\catcode `\^12\catcode `\_12\catcode `\%12\relax}%
\providecommand \@@startlink[1]{}%
\providecommand \@@endlink[0]{}%
\providecommand \url  [0]{\begingroup\@sanitize@url \@url }%
\providecommand \@url [1]{\endgroup\@href {#1}{\urlprefix }}%
\providecommand \urlprefix  [0]{URL }%
\providecommand \Eprint [0]{\href }%
\providecommand \doibase [0]{http://dx.doi.org/}%
\providecommand \selectlanguage [0]{\@gobble}%
\providecommand \bibinfo  [0]{\@secondoftwo}%
\providecommand \bibfield  [0]{\@secondoftwo}%
\providecommand \translation [1]{[#1]}%
\providecommand \BibitemOpen [0]{}%
\providecommand \bibitemStop [0]{}%
\providecommand \bibitemNoStop [0]{.\EOS\space}%
\providecommand \EOS [0]{\spacefactor3000\relax}%
\providecommand \BibitemShut  [1]{\csname bibitem#1\endcsname}%
\let\auto@bib@innerbib\@empty
\bibitem [{\citenamefont {Cheng}\ \emph {et~al.}(2006)\citenamefont {Cheng},
  \citenamefont {Low},\ and\ \citenamefont {Wang}}]{Cheng:2005as}%
  \BibitemOpen
  \bibfield  {author} {\bibinfo {author} {\bibfnamefont {H.-C.}\ \bibnamefont
  {Cheng}}, \bibinfo {author} {\bibfnamefont {I.}~\bibnamefont {Low}}, \ and\
  \bibinfo {author} {\bibfnamefont {L.-T.}\ \bibnamefont {Wang}},\ }\href
  {\doibase 10.1103/PhysRevD.74.055001} {\bibfield  {journal} {\bibinfo
  {journal} {Phys.Rev.}\ }\textbf {\bibinfo {volume} {D74}},\ \bibinfo {pages}
  {055001} (\bibinfo {year} {2006})},\ \Eprint
  {http://arxiv.org/abs/hep-ph/0510225} {arXiv:hep-ph/0510225 [hep-ph]}
  \BibitemShut {NoStop}%
\bibitem [{\citenamefont {Barger}\ \emph {et~al.}(2008)\citenamefont {Barger},
  \citenamefont {Han},\ and\ \citenamefont {Walker}}]{Barger:2006hm}%
  \BibitemOpen
  \bibfield  {author} {\bibinfo {author} {\bibfnamefont {V.}~\bibnamefont
  {Barger}}, \bibinfo {author} {\bibfnamefont {T.}~\bibnamefont {Han}}, \ and\
  \bibinfo {author} {\bibfnamefont {D.~G.}\ \bibnamefont {Walker}},\ }\href
  {\doibase 10.1103/PhysRevLett.100.031801} {\bibfield  {journal} {\bibinfo
  {journal} {Phys.Rev.Lett.}\ }\textbf {\bibinfo {volume} {100}},\ \bibinfo
  {pages} {031801} (\bibinfo {year} {2008})},\ \Eprint
  {http://arxiv.org/abs/hep-ph/0612016} {arXiv:hep-ph/0612016 [hep-ph]}
  \BibitemShut {NoStop}%
\bibitem [{\citenamefont {Agashe}\ \emph {et~al.}(2008)\citenamefont {Agashe},
  \citenamefont {Belyaev}, \citenamefont {Krupovnickas}, \citenamefont
  {Perez},\ and\ \citenamefont {Virzi}}]{Agashe:2006hk}%
  \BibitemOpen
  \bibfield  {author} {\bibinfo {author} {\bibfnamefont {K.}~\bibnamefont
  {Agashe}}, \bibinfo {author} {\bibfnamefont {A.}~\bibnamefont {Belyaev}},
  \bibinfo {author} {\bibfnamefont {T.}~\bibnamefont {Krupovnickas}}, \bibinfo
  {author} {\bibfnamefont {G.}~\bibnamefont {Perez}}, \ and\ \bibinfo {author}
  {\bibfnamefont {J.}~\bibnamefont {Virzi}},\ }\href {\doibase
  10.1103/PhysRevD.77.015003} {\bibfield  {journal} {\bibinfo  {journal}
  {Phys.Rev.}\ }\textbf {\bibinfo {volume} {D77}},\ \bibinfo {pages} {015003}
  (\bibinfo {year} {2008})},\ \Eprint {http://arxiv.org/abs/hep-ph/0612015}
  {arXiv:hep-ph/0612015 [hep-ph]} \BibitemShut {NoStop}%
\bibitem [{\citenamefont {Lillie}\ \emph {et~al.}(2007)\citenamefont {Lillie},
  \citenamefont {Randall},\ and\ \citenamefont {Wang}}]{Lillie:2007yh}%
  \BibitemOpen
  \bibfield  {author} {\bibinfo {author} {\bibfnamefont {B.}~\bibnamefont
  {Lillie}}, \bibinfo {author} {\bibfnamefont {L.}~\bibnamefont {Randall}}, \
  and\ \bibinfo {author} {\bibfnamefont {L.-T.}\ \bibnamefont {Wang}},\ }\href
  {\doibase 10.1088/1126-6708/2007/09/074} {\bibfield  {journal} {\bibinfo
  {journal} {JHEP}\ }\textbf {\bibinfo {volume} {0709}},\ \bibinfo {pages}
  {074} (\bibinfo {year} {2007})},\ \Eprint
  {http://arxiv.org/abs/hep-ph/0701166} {arXiv:hep-ph/0701166 [hep-ph]}
  \BibitemShut {NoStop}%
\bibitem [{\citenamefont {Baur}\ and\ \citenamefont {Orr}(2007)}]{Baur:2007ck}%
  \BibitemOpen
  \bibfield  {author} {\bibinfo {author} {\bibfnamefont {U.}~\bibnamefont
  {Baur}}\ and\ \bibinfo {author} {\bibfnamefont {L.}~\bibnamefont {Orr}},\
  }\href {\doibase 10.1103/PhysRevD.76.094012} {\bibfield  {journal} {\bibinfo
  {journal} {Phys.Rev.}\ }\textbf {\bibinfo {volume} {D76}},\ \bibinfo {pages}
  {094012} (\bibinfo {year} {2007})},\ \Eprint {http://arxiv.org/abs/0707.2066}
  {arXiv:0707.2066 [hep-ph]} \BibitemShut {NoStop}%
\bibitem [{\citenamefont {Baur}\ and\ \citenamefont {Orr}(2008)}]{Baur:2008uv}%
  \BibitemOpen
  \bibfield  {author} {\bibinfo {author} {\bibfnamefont {U.}~\bibnamefont
  {Baur}}\ and\ \bibinfo {author} {\bibfnamefont {L.}~\bibnamefont {Orr}},\
  }\href {\doibase 10.1103/PhysRevD.77.114001} {\bibfield  {journal} {\bibinfo
  {journal} {Phys.Rev.}\ }\textbf {\bibinfo {volume} {D77}},\ \bibinfo {pages}
  {114001} (\bibinfo {year} {2008})},\ \Eprint {http://arxiv.org/abs/0803.1160}
  {arXiv:0803.1160 [hep-ph]} \BibitemShut {NoStop}%
\bibitem [{\citenamefont {Pilot}(2013)}]{Pilot:2013bla}%
  \BibitemOpen
  \bibfield  {author} {\bibinfo {author} {\bibfnamefont {J.}~\bibnamefont
  {Pilot}} (\bibinfo {collaboration} {ATLAS, CMS Collaboration}),\ }\href
  {\doibase 10.1051/epjconf/20136009003} {\bibfield  {journal} {\bibinfo
  {journal} {EPJ Web Conf.}\ }\textbf {\bibinfo {volume} {60}},\ \bibinfo
  {pages} {09003} (\bibinfo {year} {2013})}\BibitemShut {NoStop}%
\bibitem [{\citenamefont {Fleischmann}(2013)}]{Fleischmann:2013woa}%
  \BibitemOpen
  \bibfield  {author} {\bibinfo {author} {\bibfnamefont {S.}~\bibnamefont
  {Fleischmann}} (\bibinfo {collaboration} {ATLAS, CMS Collaboration}),\ }\href
  {\doibase 10.1088/1742-6596/452/1/012034} {\bibfield  {journal} {\bibinfo
  {journal} {J.Phys.Conf.Ser.}\ }\textbf {\bibinfo {volume} {452}},\ \bibinfo
  {pages} {012034} (\bibinfo {year} {2013})}\BibitemShut {NoStop}%
\bibitem [{\citenamefont {Kaplan}\ \emph {et~al.}(2008)\citenamefont {Kaplan},
  \citenamefont {Rehermann}, \citenamefont {Schwartz},\ and\ \citenamefont
  {Tweedie}}]{Kaplan:2008ie}%
  \BibitemOpen
  \bibfield  {author} {\bibinfo {author} {\bibfnamefont {D.~E.}\ \bibnamefont
  {Kaplan}}, \bibinfo {author} {\bibfnamefont {K.}~\bibnamefont {Rehermann}},
  \bibinfo {author} {\bibfnamefont {M.~D.}\ \bibnamefont {Schwartz}}, \ and\
  \bibinfo {author} {\bibfnamefont {B.}~\bibnamefont {Tweedie}},\ }\href
  {\doibase 10.1103/PhysRevLett.101.142001} {\bibfield  {journal} {\bibinfo
  {journal} {Phys.Rev.Lett.}\ }\textbf {\bibinfo {volume} {101}},\ \bibinfo
  {pages} {142001} (\bibinfo {year} {2008})},\ \Eprint
  {http://arxiv.org/abs/0806.0848} {arXiv:0806.0848 [hep-ph]} \BibitemShut
  {NoStop}%
\bibitem [{\citenamefont {Thaler}\ and\ \citenamefont
  {Wang}(2008)}]{Thaler:2008ju}%
  \BibitemOpen
  \bibfield  {author} {\bibinfo {author} {\bibfnamefont {J.}~\bibnamefont
  {Thaler}}\ and\ \bibinfo {author} {\bibfnamefont {L.-T.}\ \bibnamefont
  {Wang}},\ }\href {\doibase 10.1088/1126-6708/2008/07/092} {\bibfield
  {journal} {\bibinfo  {journal} {JHEP}\ }\textbf {\bibinfo {volume} {0807}},\
  \bibinfo {pages} {092} (\bibinfo {year} {2008})},\ \Eprint
  {http://arxiv.org/abs/0806.0023} {arXiv:0806.0023 [hep-ph]} \BibitemShut
  {NoStop}%
\bibitem [{\citenamefont {Almeida}\ \emph
  {et~al.}(2009{\natexlab{a}})\citenamefont {Almeida}, \citenamefont {Lee},
  \citenamefont {Perez}, \citenamefont {Sterman}, \citenamefont {Sung} \emph
  {et~al.}}]{Almeida:2008yp}%
  \BibitemOpen
  \bibfield  {author} {\bibinfo {author} {\bibfnamefont {L.~G.}\ \bibnamefont
  {Almeida}}, \bibinfo {author} {\bibfnamefont {S.~J.}\ \bibnamefont {Lee}},
  \bibinfo {author} {\bibfnamefont {G.}~\bibnamefont {Perez}}, \bibinfo
  {author} {\bibfnamefont {G.~F.}\ \bibnamefont {Sterman}}, \bibinfo {author}
  {\bibfnamefont {I.}~\bibnamefont {Sung}},  \emph {et~al.},\ }\href {\doibase
  10.1103/PhysRevD.79.074017} {\bibfield  {journal} {\bibinfo  {journal}
  {Phys.Rev.}\ }\textbf {\bibinfo {volume} {D79}},\ \bibinfo {pages} {074017}
  (\bibinfo {year} {2009}{\natexlab{a}})},\ \Eprint
  {http://arxiv.org/abs/0807.0234} {arXiv:0807.0234 [hep-ph]} \BibitemShut
  {NoStop}%
\bibitem [{\citenamefont {Almeida}\ \emph
  {et~al.}(2009{\natexlab{b}})\citenamefont {Almeida}, \citenamefont {Lee},
  \citenamefont {Perez}, \citenamefont {Sung},\ and\ \citenamefont
  {Virzi}}]{Almeida:2008tp}%
  \BibitemOpen
  \bibfield  {author} {\bibinfo {author} {\bibfnamefont {L.~G.}\ \bibnamefont
  {Almeida}}, \bibinfo {author} {\bibfnamefont {S.~J.}\ \bibnamefont {Lee}},
  \bibinfo {author} {\bibfnamefont {G.}~\bibnamefont {Perez}}, \bibinfo
  {author} {\bibfnamefont {I.}~\bibnamefont {Sung}}, \ and\ \bibinfo {author}
  {\bibfnamefont {J.}~\bibnamefont {Virzi}},\ }\href {\doibase
  10.1103/PhysRevD.79.074012} {\bibfield  {journal} {\bibinfo  {journal}
  {Phys.Rev.}\ }\textbf {\bibinfo {volume} {D79}},\ \bibinfo {pages} {074012}
  (\bibinfo {year} {2009}{\natexlab{b}})},\ \Eprint
  {http://arxiv.org/abs/0810.0934} {arXiv:0810.0934 [hep-ph]} \BibitemShut
  {NoStop}%
\bibitem [{\citenamefont {Plehn}\ \emph
  {et~al.}(2010{\natexlab{a}})\citenamefont {Plehn}, \citenamefont {Salam},\
  and\ \citenamefont {Spannowsky}}]{Plehn:2009rk}%
  \BibitemOpen
  \bibfield  {author} {\bibinfo {author} {\bibfnamefont {T.}~\bibnamefont
  {Plehn}}, \bibinfo {author} {\bibfnamefont {G.~P.}\ \bibnamefont {Salam}}, \
  and\ \bibinfo {author} {\bibfnamefont {M.}~\bibnamefont {Spannowsky}},\
  }\href {\doibase 10.1103/PhysRevLett.104.111801} {\bibfield  {journal}
  {\bibinfo  {journal} {Phys.Rev.Lett.}\ }\textbf {\bibinfo {volume} {104}},\
  \bibinfo {pages} {111801} (\bibinfo {year} {2010}{\natexlab{a}})},\ \Eprint
  {http://arxiv.org/abs/0910.5472} {arXiv:0910.5472 [hep-ph]} \BibitemShut
  {NoStop}%
\bibitem [{\citenamefont {Plehn}\ \emph
  {et~al.}(2010{\natexlab{b}})\citenamefont {Plehn}, \citenamefont
  {Spannowsky}, \citenamefont {Takeuchi},\ and\ \citenamefont
  {Zerwas}}]{Plehn:2010st}%
  \BibitemOpen
  \bibfield  {author} {\bibinfo {author} {\bibfnamefont {T.}~\bibnamefont
  {Plehn}}, \bibinfo {author} {\bibfnamefont {M.}~\bibnamefont {Spannowsky}},
  \bibinfo {author} {\bibfnamefont {M.}~\bibnamefont {Takeuchi}}, \ and\
  \bibinfo {author} {\bibfnamefont {D.}~\bibnamefont {Zerwas}},\ }\href
  {\doibase 10.1007/JHEP10(2010)078} {\bibfield  {journal} {\bibinfo  {journal}
  {JHEP}\ }\textbf {\bibinfo {volume} {1010}},\ \bibinfo {pages} {078}
  (\bibinfo {year} {2010}{\natexlab{b}})},\ \Eprint
  {http://arxiv.org/abs/1006.2833} {arXiv:1006.2833 [hep-ph]} \BibitemShut
  {NoStop}%
\bibitem [{\citenamefont {Almeida}\ \emph {et~al.}(2010)\citenamefont
  {Almeida}, \citenamefont {Lee}, \citenamefont {Perez}, \citenamefont
  {Sterman},\ and\ \citenamefont {Sung}}]{Almeida:2010pa}%
  \BibitemOpen
  \bibfield  {author} {\bibinfo {author} {\bibfnamefont {L.~G.}\ \bibnamefont
  {Almeida}}, \bibinfo {author} {\bibfnamefont {S.~J.}\ \bibnamefont {Lee}},
  \bibinfo {author} {\bibfnamefont {G.}~\bibnamefont {Perez}}, \bibinfo
  {author} {\bibfnamefont {G.}~\bibnamefont {Sterman}}, \ and\ \bibinfo
  {author} {\bibfnamefont {I.}~\bibnamefont {Sung}},\ }\href {\doibase
  10.1103/PhysRevD.82.054034} {\bibfield  {journal} {\bibinfo  {journal}
  {Phys.Rev.}\ }\textbf {\bibinfo {volume} {D82}},\ \bibinfo {pages} {054034}
  (\bibinfo {year} {2010})},\ \Eprint {http://arxiv.org/abs/1006.2035}
  {arXiv:1006.2035 [hep-ph]} \BibitemShut {NoStop}%
\bibitem [{\citenamefont {Thaler}\ and\ \citenamefont
  {Van~Tilburg}(2011)}]{Thaler:2010tr}%
  \BibitemOpen
  \bibfield  {author} {\bibinfo {author} {\bibfnamefont {J.}~\bibnamefont
  {Thaler}}\ and\ \bibinfo {author} {\bibfnamefont {K.}~\bibnamefont
  {Van~Tilburg}},\ }\href {\doibase 10.1007/JHEP03(2011)015} {\bibfield
  {journal} {\bibinfo  {journal} {JHEP}\ }\textbf {\bibinfo {volume} {1103}},\
  \bibinfo {pages} {015} (\bibinfo {year} {2011})},\ \Eprint
  {http://arxiv.org/abs/1011.2268} {arXiv:1011.2268 [hep-ph]} \BibitemShut
  {NoStop}%
\bibitem [{\citenamefont {Thaler}\ and\ \citenamefont
  {Van~Tilburg}(2012)}]{Thaler:2011gf}%
  \BibitemOpen
  \bibfield  {author} {\bibinfo {author} {\bibfnamefont {J.}~\bibnamefont
  {Thaler}}\ and\ \bibinfo {author} {\bibfnamefont {K.}~\bibnamefont
  {Van~Tilburg}},\ }\href {\doibase 10.1007/JHEP02(2012)093} {\bibfield
  {journal} {\bibinfo  {journal} {JHEP}\ }\textbf {\bibinfo {volume} {1202}},\
  \bibinfo {pages} {093} (\bibinfo {year} {2012})},\ \Eprint
  {http://arxiv.org/abs/1108.2701} {arXiv:1108.2701 [hep-ph]} \BibitemShut
  {NoStop}%
\bibitem [{\citenamefont {Jankowiak}\ and\ \citenamefont
  {Larkoski}(2011)}]{Jankowiak:2011qa}%
  \BibitemOpen
  \bibfield  {author} {\bibinfo {author} {\bibfnamefont {M.}~\bibnamefont
  {Jankowiak}}\ and\ \bibinfo {author} {\bibfnamefont {A.~J.}\ \bibnamefont
  {Larkoski}},\ }\href {\doibase 10.1007/JHEP06(2011)057} {\bibfield  {journal}
  {\bibinfo  {journal} {JHEP}\ }\textbf {\bibinfo {volume} {1106}},\ \bibinfo
  {pages} {057} (\bibinfo {year} {2011})},\ \Eprint
  {http://arxiv.org/abs/1104.1646} {arXiv:1104.1646 [hep-ph]} \BibitemShut
  {NoStop}%
\bibitem [{\citenamefont {Soper}\ and\ \citenamefont
  {Spannowsky}(2013)}]{Soper:2012pb}%
  \BibitemOpen
  \bibfield  {author} {\bibinfo {author} {\bibfnamefont {D.~E.}\ \bibnamefont
  {Soper}}\ and\ \bibinfo {author} {\bibfnamefont {M.}~\bibnamefont
  {Spannowsky}},\ }\href {\doibase 10.1103/PhysRevD.87.054012} {\bibfield
  {journal} {\bibinfo  {journal} {Phys.Rev.}\ }\textbf {\bibinfo {volume}
  {D87}},\ \bibinfo {pages} {054012} (\bibinfo {year} {2013})},\ \Eprint
  {http://arxiv.org/abs/1211.3140} {arXiv:1211.3140 [hep-ph]} \BibitemShut
  {NoStop}%
\bibitem [{\citenamefont {Larkoski}\ \emph {et~al.}(2013)\citenamefont
  {Larkoski}, \citenamefont {Salam},\ and\ \citenamefont
  {Thaler}}]{Larkoski:2013eya}%
  \BibitemOpen
  \bibfield  {author} {\bibinfo {author} {\bibfnamefont {A.~J.}\ \bibnamefont
  {Larkoski}}, \bibinfo {author} {\bibfnamefont {G.~P.}\ \bibnamefont {Salam}},
  \ and\ \bibinfo {author} {\bibfnamefont {J.}~\bibnamefont {Thaler}},\ }\href
  {\doibase 10.1007/JHEP06(2013)108} {\bibfield  {journal} {\bibinfo  {journal}
  {JHEP}\ }\textbf {\bibinfo {volume} {1306}},\ \bibinfo {pages} {108}
  (\bibinfo {year} {2013})},\ \Eprint {http://arxiv.org/abs/1305.0007}
  {arXiv:1305.0007 [hep-ph]} \BibitemShut {NoStop}%
\bibitem [{925185()}]{CMS:2011bqa}%
  \BibitemOpen
  \bibfield  {author} {925185,\ }\href@noop {} {\emph {\bibinfo {title}
  {{Search for BSM ttbar Production in the Boosted All-Hadronic Final
  State}}}},\ \bibinfo {type} {Tech. Rep.}\ \bibinfo {number}
  {CMS-PAS-EXO-11-006}\ (\bibinfo {year} {2011})\BibitemShut {NoStop}%
\bibitem [{\citenamefont {Aad}\ \emph {et~al.}(2012)\citenamefont {Aad} \emph
  {et~al.}}]{Aad:2012dpa}%
  \BibitemOpen
  \bibfield  {author} {\bibinfo {author} {\bibfnamefont {G.}~\bibnamefont
  {Aad}} \emph {et~al.} (\bibinfo {collaboration} {ATLAS Collaboration}),\
  }\href {\doibase 10.1007/JHEP09(2012)041} {\bibfield  {journal} {\bibinfo
  {journal} {JHEP}\ }\textbf {\bibinfo {volume} {1209}},\ \bibinfo {pages}
  {041} (\bibinfo {year} {2012})},\ \Eprint {http://arxiv.org/abs/1207.2409}
  {arXiv:1207.2409 [hep-ex]} \BibitemShut {NoStop}%
\bibitem [{\citenamefont {Aad}\ \emph {et~al.}(2013{\natexlab{a}})\citenamefont
  {Aad} \emph {et~al.}}]{Aad:2012raa}%
  \BibitemOpen
  \bibfield  {author} {\bibinfo {author} {\bibfnamefont {G.}~\bibnamefont
  {Aad}} \emph {et~al.} (\bibinfo {collaboration} {ATLAS Collaboration}),\
  }\href {\doibase 10.1007/JHEP01(2013)116} {\bibfield  {journal} {\bibinfo
  {journal} {JHEP}\ }\textbf {\bibinfo {volume} {1301}},\ \bibinfo {pages}
  {116} (\bibinfo {year} {2013}{\natexlab{a}})},\ \Eprint
  {http://arxiv.org/abs/1211.2202} {arXiv:1211.2202 [hep-ex]} \BibitemShut
  {NoStop}%
\bibitem [{\citenamefont {Chatrchyan}\ \emph {et~al.}(2012)\citenamefont
  {Chatrchyan} \emph {et~al.}}]{Chatrchyan:2012ku}%
  \BibitemOpen
  \bibfield  {author} {\bibinfo {author} {\bibfnamefont {S.}~\bibnamefont
  {Chatrchyan}} \emph {et~al.} (\bibinfo {collaboration} {CMS Collaboration}),\
  }\href {\doibase 10.1007/JHEP09(2012)029} {\bibfield  {journal} {\bibinfo
  {journal} {JHEP}\ }\textbf {\bibinfo {volume} {1209}},\ \bibinfo {pages}
  {029} (\bibinfo {year} {2012})},\ \Eprint {http://arxiv.org/abs/1204.2488}
  {arXiv:1204.2488 [hep-ex]} \BibitemShut {NoStop}%
\bibitem [{The(2013)}]{TheATLAScollaboration:2013qia}%
  \BibitemOpen
  \href@noop {} {\emph {\bibinfo {title} {{Performance of boosted top quark
  identification in 2012 ATLAS data}}}},\ \bibinfo {type} {Tech. Rep.}\
  \bibinfo {number} {ATLAS-CONF-2013-084, ATLAS-COM-CONF-2013-074}\ (\bibinfo
  {year} {2013})\BibitemShut {NoStop}%
\bibitem [{\citenamefont {Aad}\ \emph {et~al.}(2013{\natexlab{b}})\citenamefont
  {Aad} \emph {et~al.}}]{Aad:2013gja}%
  \BibitemOpen
  \bibfield  {author} {\bibinfo {author} {\bibfnamefont {G.}~\bibnamefont
  {Aad}} \emph {et~al.} (\bibinfo {collaboration} {ATLAS Collaboration}),\
  }\href {\doibase 10.1007/JHEP09(2013)076} {\bibfield  {journal} {\bibinfo
  {journal} {JHEP}\ }\textbf {\bibinfo {volume} {1309}},\ \bibinfo {pages}
  {076} (\bibinfo {year} {2013}{\natexlab{b}})},\ \Eprint
  {http://arxiv.org/abs/1306.4945} {arXiv:1306.4945 [hep-ex]} \BibitemShut
  {NoStop}%
\bibitem [{CMS(2014)}]{CMS:2014aka}%
  \BibitemOpen
  \href@noop {} {\emph {\bibinfo {title} {{Search for top-Higgs resonances in
  all-hadronic final states using jet substructure methods}}}},\ \bibinfo
  {type} {Tech. Rep.}\ \bibinfo {number} {CMS-PAS-B2G-14-002}\ (\bibinfo {year}
  {2014})\BibitemShut {NoStop}%
\bibitem [{\citenamefont {Catani}\ \emph {et~al.}(1991)\citenamefont {Catani},
  \citenamefont {Turnock},\ and\ \citenamefont {Webber}}]{Catani:1991bd}%
  \BibitemOpen
  \bibfield  {author} {\bibinfo {author} {\bibfnamefont {S.}~\bibnamefont
  {Catani}}, \bibinfo {author} {\bibfnamefont {G.}~\bibnamefont {Turnock}}, \
  and\ \bibinfo {author} {\bibfnamefont {B.}~\bibnamefont {Webber}},\ }\href
  {\doibase 10.1016/0370-2693(91)91845-M} {\bibfield  {journal} {\bibinfo
  {journal} {Phys.Lett.}\ }\textbf {\bibinfo {volume} {B272}},\ \bibinfo
  {pages} {368} (\bibinfo {year} {1991})}\BibitemShut {NoStop}%
\bibitem [{\citenamefont {Seymour}(1998)}]{Seymour:1997kj}%
  \BibitemOpen
  \bibfield  {author} {\bibinfo {author} {\bibfnamefont {M.}~\bibnamefont
  {Seymour}},\ }\href {\doibase 10.1016/S0550-3213(97)00711-6} {\bibfield
  {journal} {\bibinfo  {journal} {Nucl.Phys.}\ }\textbf {\bibinfo {volume}
  {B513}},\ \bibinfo {pages} {269} (\bibinfo {year} {1998})},\ \Eprint
  {http://arxiv.org/abs/hep-ph/9707338} {arXiv:hep-ph/9707338 [hep-ph]}
  \BibitemShut {NoStop}%
\bibitem [{\citenamefont {Ellis}\ \emph {et~al.}(2010)\citenamefont {Ellis},
  \citenamefont {Vermilion}, \citenamefont {Walsh}, \citenamefont {Hornig},\
  and\ \citenamefont {Lee}}]{Ellis:2010rwa}%
  \BibitemOpen
  \bibfield  {author} {\bibinfo {author} {\bibfnamefont {S.~D.}\ \bibnamefont
  {Ellis}}, \bibinfo {author} {\bibfnamefont {C.~K.}\ \bibnamefont
  {Vermilion}}, \bibinfo {author} {\bibfnamefont {J.~R.}\ \bibnamefont
  {Walsh}}, \bibinfo {author} {\bibfnamefont {A.}~\bibnamefont {Hornig}}, \
  and\ \bibinfo {author} {\bibfnamefont {C.}~\bibnamefont {Lee}},\ }\href
  {\doibase 10.1007/JHEP11(2010)101} {\bibfield  {journal} {\bibinfo  {journal}
  {JHEP}\ }\textbf {\bibinfo {volume} {1011}},\ \bibinfo {pages} {101}
  (\bibinfo {year} {2010})},\ \Eprint {http://arxiv.org/abs/1001.0014}
  {arXiv:1001.0014 [hep-ph]} \BibitemShut {NoStop}%
\bibitem [{\citenamefont {Chien}\ and\ \citenamefont
  {Schwartz}(2010)}]{Chien:2010kc}%
  \BibitemOpen
  \bibfield  {author} {\bibinfo {author} {\bibfnamefont {Y.-T.}\ \bibnamefont
  {Chien}}\ and\ \bibinfo {author} {\bibfnamefont {M.~D.}\ \bibnamefont
  {Schwartz}},\ }\href {\doibase 10.1007/JHEP08(2010)058} {\bibfield  {journal}
  {\bibinfo  {journal} {JHEP}\ }\textbf {\bibinfo {volume} {1008}},\ \bibinfo
  {pages} {058} (\bibinfo {year} {2010})},\ \Eprint
  {http://arxiv.org/abs/1005.1644} {arXiv:1005.1644 [hep-ph]} \BibitemShut
  {NoStop}%
\bibitem [{\citenamefont {Feige}\ \emph {et~al.}(2012)\citenamefont {Feige},
  \citenamefont {Schwartz}, \citenamefont {Stewart},\ and\ \citenamefont
  {Thaler}}]{Feige:2012vc}%
  \BibitemOpen
  \bibfield  {author} {\bibinfo {author} {\bibfnamefont {I.}~\bibnamefont
  {Feige}}, \bibinfo {author} {\bibfnamefont {M.~D.}\ \bibnamefont {Schwartz}},
  \bibinfo {author} {\bibfnamefont {I.~W.}\ \bibnamefont {Stewart}}, \ and\
  \bibinfo {author} {\bibfnamefont {J.}~\bibnamefont {Thaler}},\ }\href
  {\doibase 10.1103/PhysRevLett.109.092001} {\bibfield  {journal} {\bibinfo
  {journal} {Phys.Rev.Lett.}\ }\textbf {\bibinfo {volume} {109}},\ \bibinfo
  {pages} {092001} (\bibinfo {year} {2012})},\ \Eprint
  {http://arxiv.org/abs/1204.3898} {arXiv:1204.3898 [hep-ph]} \BibitemShut
  {NoStop}%
\bibitem [{\citenamefont {Dasgupta}\ \emph {et~al.}(2012)\citenamefont
  {Dasgupta}, \citenamefont {Khelifa-Kerfa}, \citenamefont {Marzani},\ and\
  \citenamefont {Spannowsky}}]{Dasgupta:2012hg}%
  \BibitemOpen
  \bibfield  {author} {\bibinfo {author} {\bibfnamefont {M.}~\bibnamefont
  {Dasgupta}}, \bibinfo {author} {\bibfnamefont {K.}~\bibnamefont
  {Khelifa-Kerfa}}, \bibinfo {author} {\bibfnamefont {S.}~\bibnamefont
  {Marzani}}, \ and\ \bibinfo {author} {\bibfnamefont {M.}~\bibnamefont
  {Spannowsky}},\ }\href {\doibase 10.1007/JHEP10(2012)126} {\bibfield
  {journal} {\bibinfo  {journal} {JHEP}\ }\textbf {\bibinfo {volume} {1210}},\
  \bibinfo {pages} {126} (\bibinfo {year} {2012})},\ \Eprint
  {http://arxiv.org/abs/1207.1640} {arXiv:1207.1640 [hep-ph]} \BibitemShut
  {NoStop}%
\bibitem [{\citenamefont {Chien}\ \emph {et~al.}(2013)\citenamefont {Chien},
  \citenamefont {Kelley}, \citenamefont {Schwartz},\ and\ \citenamefont
  {Zhu}}]{Chien:2012ur}%
  \BibitemOpen
  \bibfield  {author} {\bibinfo {author} {\bibfnamefont {Y.-T.}\ \bibnamefont
  {Chien}}, \bibinfo {author} {\bibfnamefont {R.}~\bibnamefont {Kelley}},
  \bibinfo {author} {\bibfnamefont {M.~D.}\ \bibnamefont {Schwartz}}, \ and\
  \bibinfo {author} {\bibfnamefont {H.~X.}\ \bibnamefont {Zhu}},\ }\href
  {\doibase 10.1103/PhysRevD.87.014010} {\bibfield  {journal} {\bibinfo
  {journal} {Phys.Rev.}\ }\textbf {\bibinfo {volume} {D87}},\ \bibinfo {pages}
  {014010} (\bibinfo {year} {2013})},\ \Eprint {http://arxiv.org/abs/1208.0010}
  {arXiv:1208.0010} \BibitemShut {NoStop}%
\bibitem [{\citenamefont {Krohn}\ \emph {et~al.}(2013)\citenamefont {Krohn},
  \citenamefont {Schwartz}, \citenamefont {Lin},\ and\ \citenamefont
  {Waalewijn}}]{Krohn:2012fg}%
  \BibitemOpen
  \bibfield  {author} {\bibinfo {author} {\bibfnamefont {D.}~\bibnamefont
  {Krohn}}, \bibinfo {author} {\bibfnamefont {M.~D.}\ \bibnamefont {Schwartz}},
  \bibinfo {author} {\bibfnamefont {T.}~\bibnamefont {Lin}}, \ and\ \bibinfo
  {author} {\bibfnamefont {W.~J.}\ \bibnamefont {Waalewijn}},\ }\href {\doibase
  10.1103/PhysRevLett.110.212001} {\bibfield  {journal} {\bibinfo  {journal}
  {Phys.Rev.Lett.}\ }\textbf {\bibinfo {volume} {110}},\ \bibinfo {pages}
  {212001} (\bibinfo {year} {2013})},\ \Eprint {http://arxiv.org/abs/1209.2421}
  {arXiv:1209.2421 [hep-ph]} \BibitemShut {NoStop}%
\bibitem [{\citenamefont {Waalewijn}(2012)}]{Waalewijn:2012sv}%
  \BibitemOpen
  \bibfield  {author} {\bibinfo {author} {\bibfnamefont {W.~J.}\ \bibnamefont
  {Waalewijn}},\ }\href {\doibase 10.1103/PhysRevD.86.094030} {\bibfield
  {journal} {\bibinfo  {journal} {Phys.Rev.}\ }\textbf {\bibinfo {volume}
  {D86}},\ \bibinfo {pages} {094030} (\bibinfo {year} {2012})},\ \Eprint
  {http://arxiv.org/abs/1209.3019} {arXiv:1209.3019 [hep-ph]} \BibitemShut
  {NoStop}%
\bibitem [{\citenamefont {Field}\ \emph {et~al.}(2013)\citenamefont {Field},
  \citenamefont {Gur-Ari}, \citenamefont {Kosower}, \citenamefont {Mannelli},\
  and\ \citenamefont {Perez}}]{Field:2012rw}%
  \BibitemOpen
  \bibfield  {author} {\bibinfo {author} {\bibfnamefont {M.}~\bibnamefont
  {Field}}, \bibinfo {author} {\bibfnamefont {G.}~\bibnamefont {Gur-Ari}},
  \bibinfo {author} {\bibfnamefont {D.~A.}\ \bibnamefont {Kosower}}, \bibinfo
  {author} {\bibfnamefont {L.}~\bibnamefont {Mannelli}}, \ and\ \bibinfo
  {author} {\bibfnamefont {G.}~\bibnamefont {Perez}},\ }\href {\doibase
  10.1103/PhysRevD.87.094013} {\bibfield  {journal} {\bibinfo  {journal}
  {Phys.Rev.}\ }\textbf {\bibinfo {volume} {D87}},\ \bibinfo {pages} {094013}
  (\bibinfo {year} {2013})},\ \Eprint {http://arxiv.org/abs/1212.2106}
  {arXiv:1212.2106 [hep-ph]} \BibitemShut {NoStop}%
\bibitem [{\citenamefont {Jouttenus}\ \emph {et~al.}(2013)\citenamefont
  {Jouttenus}, \citenamefont {Stewart}, \citenamefont {Tackmann},\ and\
  \citenamefont {Waalewijn}}]{Jouttenus:2013hs}%
  \BibitemOpen
  \bibfield  {author} {\bibinfo {author} {\bibfnamefont {T.~T.}\ \bibnamefont
  {Jouttenus}}, \bibinfo {author} {\bibfnamefont {I.~W.}\ \bibnamefont
  {Stewart}}, \bibinfo {author} {\bibfnamefont {F.~J.}\ \bibnamefont
  {Tackmann}}, \ and\ \bibinfo {author} {\bibfnamefont {W.~J.}\ \bibnamefont
  {Waalewijn}},\ }\href {\doibase 10.1103/PhysRevD.88.054031} {\bibfield
  {journal} {\bibinfo  {journal} {Phys.Rev.}\ }\textbf {\bibinfo {volume}
  {D88}},\ \bibinfo {pages} {054031} (\bibinfo {year} {2013})},\ \Eprint
  {http://arxiv.org/abs/1302.0846} {arXiv:1302.0846 [hep-ph]} \BibitemShut
  {NoStop}%
\bibitem [{\citenamefont {Dasgupta}\ \emph
  {et~al.}(2013{\natexlab{a}})\citenamefont {Dasgupta}, \citenamefont
  {Fregoso}, \citenamefont {Marzani},\ and\ \citenamefont
  {Powling}}]{Dasgupta:2013via}%
  \BibitemOpen
  \bibfield  {author} {\bibinfo {author} {\bibfnamefont {M.}~\bibnamefont
  {Dasgupta}}, \bibinfo {author} {\bibfnamefont {A.}~\bibnamefont {Fregoso}},
  \bibinfo {author} {\bibfnamefont {S.}~\bibnamefont {Marzani}}, \ and\
  \bibinfo {author} {\bibfnamefont {A.}~\bibnamefont {Powling}},\ }\href
  {\doibase 10.1140/epjc/s10052-013-2623-3} {\bibfield  {journal} {\bibinfo
  {journal} {Eur.Phys.J.}\ }\textbf {\bibinfo {volume} {C73}},\ \bibinfo
  {pages} {2623} (\bibinfo {year} {2013}{\natexlab{a}})},\ \Eprint
  {http://arxiv.org/abs/1307.0013} {arXiv:1307.0013 [hep-ph]} \BibitemShut
  {NoStop}%
\bibitem [{\citenamefont {Dasgupta}\ \emph
  {et~al.}(2013{\natexlab{b}})\citenamefont {Dasgupta}, \citenamefont
  {Fregoso}, \citenamefont {Marzani},\ and\ \citenamefont
  {Salam}}]{Dasgupta:2013ihk}%
  \BibitemOpen
  \bibfield  {author} {\bibinfo {author} {\bibfnamefont {M.}~\bibnamefont
  {Dasgupta}}, \bibinfo {author} {\bibfnamefont {A.}~\bibnamefont {Fregoso}},
  \bibinfo {author} {\bibfnamefont {S.}~\bibnamefont {Marzani}}, \ and\
  \bibinfo {author} {\bibfnamefont {G.~P.}\ \bibnamefont {Salam}},\ }\href
  {\doibase 10.1007/JHEP09(2013)029} {\bibfield  {journal} {\bibinfo  {journal}
  {JHEP}\ }\textbf {\bibinfo {volume} {1309}},\ \bibinfo {pages} {029}
  (\bibinfo {year} {2013}{\natexlab{b}})},\ \Eprint
  {http://arxiv.org/abs/1307.0007} {arXiv:1307.0007 [hep-ph]} \BibitemShut
  {NoStop}%
\bibitem [{\citenamefont {Larkoski}\ and\ \citenamefont
  {Thaler}(2013)}]{Larkoski:2013paa}%
  \BibitemOpen
  \bibfield  {author} {\bibinfo {author} {\bibfnamefont {A.~J.}\ \bibnamefont
  {Larkoski}}\ and\ \bibinfo {author} {\bibfnamefont {J.}~\bibnamefont
  {Thaler}},\ }\href {\doibase 10.1007/JHEP09(2013)137} {\bibfield  {journal}
  {\bibinfo  {journal} {JHEP}\ }\textbf {\bibinfo {volume} {1309}},\ \bibinfo
  {pages} {137} (\bibinfo {year} {2013})},\ \Eprint
  {http://arxiv.org/abs/1307.1699} {arXiv:1307.1699} \BibitemShut {NoStop}%
\bibitem [{\citenamefont {Larkoski}\ \emph
  {et~al.}(2014{\natexlab{a}})\citenamefont {Larkoski}, \citenamefont {Neill},\
  and\ \citenamefont {Thaler}}]{Larkoski:2014uqa}%
  \BibitemOpen
  \bibfield  {author} {\bibinfo {author} {\bibfnamefont {A.~J.}\ \bibnamefont
  {Larkoski}}, \bibinfo {author} {\bibfnamefont {D.}~\bibnamefont {Neill}}, \
  and\ \bibinfo {author} {\bibfnamefont {J.}~\bibnamefont {Thaler}},\ }\href
  {\doibase 10.1007/JHEP04(2014)017} {\bibfield  {journal} {\bibinfo  {journal}
  {JHEP}\ }\textbf {\bibinfo {volume} {1404}},\ \bibinfo {pages} {017}
  (\bibinfo {year} {2014}{\natexlab{a}})},\ \Eprint
  {http://arxiv.org/abs/1401.2158} {arXiv:1401.2158 [hep-ph]} \BibitemShut
  {NoStop}%
\bibitem [{\citenamefont {Larkoski}\ \emph
  {et~al.}(2014{\natexlab{b}})\citenamefont {Larkoski}, \citenamefont {Moult},\
  and\ \citenamefont {Neill}}]{Larkoski:2014tva}%
  \BibitemOpen
  \bibfield  {author} {\bibinfo {author} {\bibfnamefont {A.~J.}\ \bibnamefont
  {Larkoski}}, \bibinfo {author} {\bibfnamefont {I.}~\bibnamefont {Moult}}, \
  and\ \bibinfo {author} {\bibfnamefont {D.}~\bibnamefont {Neill}},\
  }\href@noop {} {\  (\bibinfo {year} {2014}{\natexlab{b}})},\ \Eprint
  {http://arxiv.org/abs/1401.4458} {arXiv:1401.4458 [hep-ph]} \BibitemShut
  {NoStop}%
\bibitem [{\citenamefont {Larkoski}\ \emph
  {et~al.}(2014{\natexlab{c}})\citenamefont {Larkoski}, \citenamefont
  {Thaler},\ and\ \citenamefont {Waalewijn}}]{Larkoski:2014pca}%
  \BibitemOpen
  \bibfield  {author} {\bibinfo {author} {\bibfnamefont {A.~J.}\ \bibnamefont
  {Larkoski}}, \bibinfo {author} {\bibfnamefont {J.}~\bibnamefont {Thaler}}, \
  and\ \bibinfo {author} {\bibfnamefont {W.~J.}\ \bibnamefont {Waalewijn}},\
  }\href@noop {} {\  (\bibinfo {year} {2014}{\natexlab{c}})},\ \Eprint
  {http://arxiv.org/abs/1408.3122} {arXiv:1408.3122 [hep-ph]} \BibitemShut
  {NoStop}%
\bibitem [{\citenamefont {Procura}\ \emph {et~al.}(2014)\citenamefont
  {Procura}, \citenamefont {Waalewijn},\ and\ \citenamefont
  {Zeune}}]{Procura:2014cba}%
  \BibitemOpen
  \bibfield  {author} {\bibinfo {author} {\bibfnamefont {M.}~\bibnamefont
  {Procura}}, \bibinfo {author} {\bibfnamefont {W.~J.}\ \bibnamefont
  {Waalewijn}}, \ and\ \bibinfo {author} {\bibfnamefont {L.}~\bibnamefont
  {Zeune}},\ }\href@noop {} {\  (\bibinfo {year} {2014})},\ \Eprint
  {http://arxiv.org/abs/1410.6483} {arXiv:1410.6483 [hep-ph]} \BibitemShut
  {NoStop}%
\bibitem [{\citenamefont {Larkoski}\ \emph {et~al.}(ming)\citenamefont
  {Larkoski}, \citenamefont {Moult},\ and\ \citenamefont {Neill}}]{usD2}%
  \BibitemOpen
  \bibfield  {author} {\bibinfo {author} {\bibfnamefont {A.~J.}\ \bibnamefont
  {Larkoski}}, \bibinfo {author} {\bibfnamefont {I.}~\bibnamefont {Moult}}, \
  and\ \bibinfo {author} {\bibfnamefont {D.}~\bibnamefont {Neill}},\
  }\href@noop {} {\bibfield  {journal} {\bibinfo  {journal} {Resumming
  phenomenological Jet Observables}\ } (\bibinfo {year}
  {Forthcoming})}\BibitemShut {NoStop}%
\bibitem [{\citenamefont {Larkoski}\ \emph
  {et~al.}(2014{\natexlab{d}})\citenamefont {Larkoski}, \citenamefont {Moult},\
  and\ \citenamefont {Neill}}]{Larkoski:2014gra}%
  \BibitemOpen
  \bibfield  {author} {\bibinfo {author} {\bibfnamefont {A.~J.}\ \bibnamefont
  {Larkoski}}, \bibinfo {author} {\bibfnamefont {I.}~\bibnamefont {Moult}}, \
  and\ \bibinfo {author} {\bibfnamefont {D.}~\bibnamefont {Neill}},\
  }\href@noop {} {\  (\bibinfo {year} {2014}{\natexlab{d}})},\ \Eprint
  {http://arxiv.org/abs/1409.6298} {arXiv:1409.6298 [hep-ph]} \BibitemShut
  {NoStop}%
\bibitem [{\citenamefont {Walsh}\ and\ \citenamefont
  {Zuberi}(2011)}]{Walsh:2011fz}%
  \BibitemOpen
  \bibfield  {author} {\bibinfo {author} {\bibfnamefont {J.~R.}\ \bibnamefont
  {Walsh}}\ and\ \bibinfo {author} {\bibfnamefont {S.}~\bibnamefont {Zuberi}},\
  }\href@noop {} {\  (\bibinfo {year} {2011})},\ \Eprint
  {http://arxiv.org/abs/1110.5333} {arXiv:1110.5333 [hep-ph]} \BibitemShut
  {NoStop}%
\bibitem [{\citenamefont {Bauer}\ \emph {et~al.}(2012)\citenamefont {Bauer},
  \citenamefont {Tackmann}, \citenamefont {Walsh},\ and\ \citenamefont
  {Zuberi}}]{Bauer:2011uc}%
  \BibitemOpen
  \bibfield  {author} {\bibinfo {author} {\bibfnamefont {C.~W.}\ \bibnamefont
  {Bauer}}, \bibinfo {author} {\bibfnamefont {F.~J.}\ \bibnamefont {Tackmann}},
  \bibinfo {author} {\bibfnamefont {J.~R.}\ \bibnamefont {Walsh}}, \ and\
  \bibinfo {author} {\bibfnamefont {S.}~\bibnamefont {Zuberi}},\ }\href
  {\doibase 10.1103/PhysRevD.85.074006} {\bibfield  {journal} {\bibinfo
  {journal} {Phys.Rev.}\ }\textbf {\bibinfo {volume} {D85}},\ \bibinfo {pages}
  {074006} (\bibinfo {year} {2012})},\ \Eprint {http://arxiv.org/abs/1106.6047}
  {arXiv:1106.6047 [hep-ph]} \BibitemShut {NoStop}%
\bibitem [{\citenamefont {Sjostrand}\ \emph {et~al.}(2006)\citenamefont
  {Sjostrand}, \citenamefont {Mrenna},\ and\ \citenamefont
  {Skands}}]{Sjostrand:2006za}%
  \BibitemOpen
  \bibfield  {author} {\bibinfo {author} {\bibfnamefont {T.}~\bibnamefont
  {Sjostrand}}, \bibinfo {author} {\bibfnamefont {S.}~\bibnamefont {Mrenna}}, \
  and\ \bibinfo {author} {\bibfnamefont {P.~Z.}\ \bibnamefont {Skands}},\
  }\href {\doibase 10.1088/1126-6708/2006/05/026} {\bibfield  {journal}
  {\bibinfo  {journal} {JHEP}\ }\textbf {\bibinfo {volume} {0605}},\ \bibinfo
  {pages} {026} (\bibinfo {year} {2006})},\ \Eprint
  {http://arxiv.org/abs/hep-ph/0603175} {arXiv:hep-ph/0603175 [hep-ph]}
  \BibitemShut {NoStop}%
\bibitem [{\citenamefont {Sjostrand}\ \emph {et~al.}(2008)\citenamefont
  {Sjostrand}, \citenamefont {Mrenna},\ and\ \citenamefont
  {Skands}}]{Sjostrand:2007gs}%
  \BibitemOpen
  \bibfield  {author} {\bibinfo {author} {\bibfnamefont {T.}~\bibnamefont
  {Sjostrand}}, \bibinfo {author} {\bibfnamefont {S.}~\bibnamefont {Mrenna}}, \
  and\ \bibinfo {author} {\bibfnamefont {P.~Z.}\ \bibnamefont {Skands}},\
  }\href {\doibase 10.1016/j.cpc.2008.01.036} {\bibfield  {journal} {\bibinfo
  {journal} {Comput.Phys.Commun.}\ }\textbf {\bibinfo {volume} {178}},\
  \bibinfo {pages} {852} (\bibinfo {year} {2008})},\ \Eprint
  {http://arxiv.org/abs/0710.3820} {arXiv:0710.3820 [hep-ph]} \BibitemShut
  {NoStop}%
\bibitem [{\citenamefont {Marchesini}\ \emph {et~al.}(1992)\citenamefont
  {Marchesini}, \citenamefont {Webber}, \citenamefont {Abbiendi}, \citenamefont
  {Knowles}, \citenamefont {Seymour} \emph {et~al.}}]{Marchesini:1991ch}%
  \BibitemOpen
  \bibfield  {author} {\bibinfo {author} {\bibfnamefont {G.}~\bibnamefont
  {Marchesini}}, \bibinfo {author} {\bibfnamefont {B.}~\bibnamefont {Webber}},
  \bibinfo {author} {\bibfnamefont {G.}~\bibnamefont {Abbiendi}}, \bibinfo
  {author} {\bibfnamefont {I.}~\bibnamefont {Knowles}}, \bibinfo {author}
  {\bibfnamefont {M.}~\bibnamefont {Seymour}},  \emph {et~al.},\ }\href
  {\doibase 10.1016/0010-4655(92)90055-4} {\bibfield  {journal} {\bibinfo
  {journal} {Comput.Phys.Commun.}\ }\textbf {\bibinfo {volume} {67}},\ \bibinfo
  {pages} {465} (\bibinfo {year} {1992})}\BibitemShut {NoStop}%
\bibitem [{\citenamefont {Corcella}\ \emph {et~al.}(2001)\citenamefont
  {Corcella}, \citenamefont {Knowles}, \citenamefont {Marchesini},
  \citenamefont {Moretti}, \citenamefont {Odagiri} \emph
  {et~al.}}]{Corcella:2000bw}%
  \BibitemOpen
  \bibfield  {author} {\bibinfo {author} {\bibfnamefont {G.}~\bibnamefont
  {Corcella}}, \bibinfo {author} {\bibfnamefont {I.}~\bibnamefont {Knowles}},
  \bibinfo {author} {\bibfnamefont {G.}~\bibnamefont {Marchesini}}, \bibinfo
  {author} {\bibfnamefont {S.}~\bibnamefont {Moretti}}, \bibinfo {author}
  {\bibfnamefont {K.}~\bibnamefont {Odagiri}},  \emph {et~al.},\ }\href
  {\doibase 10.1088/1126-6708/2001/01/010} {\bibfield  {journal} {\bibinfo
  {journal} {JHEP}\ }\textbf {\bibinfo {volume} {0101}},\ \bibinfo {pages}
  {010} (\bibinfo {year} {2001})},\ \Eprint
  {http://arxiv.org/abs/hep-ph/0011363} {arXiv:hep-ph/0011363 [hep-ph]}
  \BibitemShut {NoStop}%
\bibitem [{\citenamefont {Corcella}\ \emph {et~al.}(2002)\citenamefont
  {Corcella}, \citenamefont {Knowles}, \citenamefont {Marchesini},
  \citenamefont {Moretti}, \citenamefont {Odagiri} \emph
  {et~al.}}]{Corcella:2002jc}%
  \BibitemOpen
  \bibfield  {author} {\bibinfo {author} {\bibfnamefont {G.}~\bibnamefont
  {Corcella}}, \bibinfo {author} {\bibfnamefont {I.}~\bibnamefont {Knowles}},
  \bibinfo {author} {\bibfnamefont {G.}~\bibnamefont {Marchesini}}, \bibinfo
  {author} {\bibfnamefont {S.}~\bibnamefont {Moretti}}, \bibinfo {author}
  {\bibfnamefont {K.}~\bibnamefont {Odagiri}},  \emph {et~al.},\ }\href@noop {}
  {\  (\bibinfo {year} {2002})},\ \Eprint {http://arxiv.org/abs/hep-ph/0210213}
  {arXiv:hep-ph/0210213 [hep-ph]} \BibitemShut {NoStop}%
\bibitem [{\citenamefont {Bahr}\ \emph {et~al.}(2008)\citenamefont {Bahr},
  \citenamefont {Gieseke}, \citenamefont {Gigg}, \citenamefont {Grellscheid},
  \citenamefont {Hamilton} \emph {et~al.}}]{Bahr:2008pv}%
  \BibitemOpen
  \bibfield  {author} {\bibinfo {author} {\bibfnamefont {M.}~\bibnamefont
  {Bahr}}, \bibinfo {author} {\bibfnamefont {S.}~\bibnamefont {Gieseke}},
  \bibinfo {author} {\bibfnamefont {M.}~\bibnamefont {Gigg}}, \bibinfo {author}
  {\bibfnamefont {D.}~\bibnamefont {Grellscheid}}, \bibinfo {author}
  {\bibfnamefont {K.}~\bibnamefont {Hamilton}},  \emph {et~al.},\ }\href
  {\doibase 10.1140/epjc/s10052-008-0798-9} {\bibfield  {journal} {\bibinfo
  {journal} {Eur.Phys.J.}\ }\textbf {\bibinfo {volume} {C58}},\ \bibinfo
  {pages} {639} (\bibinfo {year} {2008})},\ \Eprint
  {http://arxiv.org/abs/0803.0883} {arXiv:0803.0883 [hep-ph]} \BibitemShut
  {NoStop}%
\bibitem [{\citenamefont {Cacciari}\ \emph {et~al.}(2012)\citenamefont
  {Cacciari}, \citenamefont {Salam},\ and\ \citenamefont
  {Soyez}}]{Cacciari:2011ma}%
  \BibitemOpen
  \bibfield  {author} {\bibinfo {author} {\bibfnamefont {M.}~\bibnamefont
  {Cacciari}}, \bibinfo {author} {\bibfnamefont {G.~P.}\ \bibnamefont {Salam}},
  \ and\ \bibinfo {author} {\bibfnamefont {G.}~\bibnamefont {Soyez}},\ }\href
  {\doibase 10.1140/epjc/s10052-012-1896-2} {\bibfield  {journal} {\bibinfo
  {journal} {Eur.Phys.J.}\ }\textbf {\bibinfo {volume} {C72}},\ \bibinfo
  {pages} {1896} (\bibinfo {year} {2012})},\ \Eprint
  {http://arxiv.org/abs/1111.6097} {arXiv:1111.6097 [hep-ph]} \BibitemShut
  {NoStop}%
\bibitem [{\citenamefont {Cacciari}\ \emph {et~al.}(2008)\citenamefont
  {Cacciari}, \citenamefont {Salam},\ and\ \citenamefont
  {Soyez}}]{Cacciari:2008gp}%
  \BibitemOpen
  \bibfield  {author} {\bibinfo {author} {\bibfnamefont {M.}~\bibnamefont
  {Cacciari}}, \bibinfo {author} {\bibfnamefont {G.~P.}\ \bibnamefont {Salam}},
  \ and\ \bibinfo {author} {\bibfnamefont {G.}~\bibnamefont {Soyez}},\ }\href
  {\doibase 10.1088/1126-6708/2008/04/063} {\bibfield  {journal} {\bibinfo
  {journal} {JHEP}\ }\textbf {\bibinfo {volume} {0804}},\ \bibinfo {pages}
  {063} (\bibinfo {year} {2008})},\ \Eprint {http://arxiv.org/abs/0802.1189}
  {arXiv:0802.1189 [hep-ph]} \BibitemShut {NoStop}%
\bibitem [{\citenamefont {Bertolini}\ \emph {et~al.}(2014)\citenamefont
  {Bertolini}, \citenamefont {Chan},\ and\ \citenamefont
  {Thaler}}]{Bertolini:2013iqa}%
  \BibitemOpen
  \bibfield  {author} {\bibinfo {author} {\bibfnamefont {D.}~\bibnamefont
  {Bertolini}}, \bibinfo {author} {\bibfnamefont {T.}~\bibnamefont {Chan}}, \
  and\ \bibinfo {author} {\bibfnamefont {J.}~\bibnamefont {Thaler}},\ }\href
  {\doibase 10.1007/JHEP04(2014)013} {\bibfield  {journal} {\bibinfo  {journal}
  {JHEP}\ }\textbf {\bibinfo {volume} {1404}},\ \bibinfo {pages} {013}
  (\bibinfo {year} {2014})},\ \Eprint {http://arxiv.org/abs/1310.7584}
  {arXiv:1310.7584 [hep-ph]} \BibitemShut {NoStop}%
\bibitem [{\citenamefont {Larkoski}\ and\ \citenamefont
  {Thaler}(2014)}]{Larkoski:2014bia}%
  \BibitemOpen
  \bibfield  {author} {\bibinfo {author} {\bibfnamefont {A.~J.}\ \bibnamefont
  {Larkoski}}\ and\ \bibinfo {author} {\bibfnamefont {J.}~\bibnamefont
  {Thaler}},\ }\href@noop {} {\  (\bibinfo {year} {2014})},\ \Eprint
  {http://arxiv.org/abs/1406.7011} {arXiv:1406.7011 [hep-ph]} \BibitemShut
  {NoStop}%
\bibitem [{\citenamefont {Salam}()}]{Salambroadening}%
  \BibitemOpen
  \bibfield  {author} {\bibinfo {author} {\bibfnamefont {G.}~\bibnamefont
  {Salam}},\ }\href@noop {} {\bibinfo  {journal} {Unpublished}\ }\BibitemShut
  {NoStop}%
\bibitem [{fjc()}]{fjcontrib}%
  \BibitemOpen
\bibfield  {journal} {  }\href {http://fastjet.hepforge.org/contrib/} {\enquote
  {\bibinfo {title} {Fastjet contrib},}\ }\bibinfo {howpublished}
  {\url{http://fastjet.hepforge.org/contrib/}}\BibitemShut {NoStop}%
\end{thebibliography}%

\end{document}